\documentclass[9pt,twocolumn,twoside]{opticajnl}
\journal{opticajournal}
\setboolean{shortarticle}{false}
\usepackage{lineno}

\usepackage{bm}
\usepackage{hyperref}
\usepackage [english]{babel}
\usepackage [utf8]{inputenc}
\usepackage [T1]{fontenc}
\usepackage{graphicx}
\usepackage{amsmath}
\usepackage{amssymb}
\usepackage{siunitx}
\usepackage{booktabs}
\usepackage{array}
\usepackage{here}
\usepackage[normalem]{ulem}
\usepackage{cuted}
\providecommand{\ifhighlighting}{\iffalse}
\ifhighlighting
\else
\newcommand{\hlstart}{}

\fi

\newcommand{\be}{\begin{equation}}
\newcommand{\e}{\end{equation}}

\newcommand{\beml}{\begin{subequations}}
\newcommand{\eml}{\end{subequations}}
\newcommand{\beq}{\begin{eqnarray}}
\newcommand{\eq}{\end{eqnarray}}
\newcommand{\ba}{\begin{array}}
\newcommand{\ea}{\end{array}}

\usepackage{braket}

\usepackage{nicefrac}
\usepackage{tikz}
\usetikzlibrary{arrows.meta}
\usetikzlibrary{patterns}
\usepackage{diagbox}

\date{\today}

\title{Accurate Zernike-Corrected Phase Screens for Arbitrary Power Spectra}

\author[1,*]{David Bachmann}
\author[1,$\dagger$]{Mathieu Isoard}
\author[1,2]{Vyacheslav Shatokhin}
\author[3]{Giacomo Sorelli}
\author[1,2]{Andreas Buchleitner}

\affil[1]{Physikalisches Institut, Albert-Ludwigs-Universit\"at Freiburg, Hermann-Herder-Str.\ 3,
79104 Freiburg, Germany}

\affil[2]{EUCOR Centre for Quantum Science and Quantum Computing, Albert-Ludwigs-Universit\"at Freiburg, Hermann-Herder-Str. 3, 79104 Freiburg, Germany}

\affil[3]{Fraunhofer IOSB, Ettlingen, Fraunhofer Institute of Optronics, System Technologies and Image Exploitation, Gutleuthausstr. 1, 76275 Ettlingen, Germany}

\affil[$\dagger$]{Current address: Laboratoire Kastler Brossel, Sorbonne Université, ENS-Université PSL, Collège de France, CNRS; 4 place Jussieu, 75252 Paris, France}

\affil[*]{david.bachmann@physik.uni-freiburg.de}

\begin{abstract}
Wave propagation through 
random continuous
media
remains an 
important fundamental
problem with applications ranging from remote sensing to quantum communication.
Typically, such media are characterized 
by smooth refractive index fluctuations whose impact
on the wave can be captured by the stochastic parabolic equation.
The latter 
can
be 
solved numerically by means of a split-step method, 
which replaces the continuous medium with a number of discrete phase screens derived from the medium's power spectrum.  
We introduce and benchmark highly accurate and efficient hybrid phase screens for arbitrary power spectra which are based on the combination of Zernike and Fourier phase screens.
\end{abstract}

\setboolean{displaycopyright}{false}

\begin{document}

\maketitle

\section{Introduction}
\label{sec:intro}
\hlstart

Wave propagation through complex stochastic 
media 
is a ubiquitous physical process.
An important class of such media, exemplified by planetary atmospheres, clear air and water turbulence, or biological tissue, are random continua 
whose index of refraction varies smoothly in space and time \cite{Ishimaru78}. 
Motivated by numerous applications including remote sensing, imaging and optical communication \cite{Roddier04, Willner21, Hufnagel20, Gigan22}, current research strives for
a more accurate quantitative description of effects
induced by such random continua on the propagation of light. 

Let us 
consider optical wave propagation through clear air or underwater turbulence, where the typical size of the continuous refractive index's inhomogeneities is much larger than the light's wavelength. Thus, mainly forward- and small-angle scattering with respect to the propagation direction takes place, that can be described by the \emph{stochastic parabolic equation} \cite{Ishimaru78,Andrews05}. Via the fluctuating part of the refractive index, the latter captures the principal effects of turbulence on the propagating wave: phase and intensity fluctuations (scintillation), beam wandering and broadening. 

Although in 
some limiting cases the stochastic parabolic equation can be solved analytically \cite{Ishimaru78}, in general one needs to resort to numerical techniques. Therefore,
developing accurate numerical models of turbulence, to predict and understand how light is affected by turbulence, is of high importance.
In this context, 
the \emph{split-step method} has proven to be efficient \cite{Schmidt10}. The main idea of the method is to replace the medium with a sequence of discrete \emph{phase screens} which mimic narrow layers of the medium with (spatially) fluctuating index of refraction, importing
phase shifts upon the wave's transverse profile.
Between the screens, the wave undergoes free diffraction.
Inherently, the accuracy of this method hinges upon the phase screens' statistically correct representation of the medium's power spectrum, whose Fourier transform yields the two-point spatial correlations of the fluctuating part of the refractive index \cite{Ishimaru78,Andrews05}.
Furthermore, phase screens with arbitrary statistics can be efficiently implemented using spatial light modulators, allowing for emulation of complex random media in table top experiments \cite{Rodenburg14,Bachmann24}.

In the past decades, several different approaches 
to
the generation of phase screens were 
developed \cite{Lane92, Mcglamery76, Martin88, Roddier90, Herman90, Johansson94, Frehlich00, Sedmak04, Asse06, Charnotskii:13,Anzuola17}.
Most prominently, spectral filtering based on the fast Fourier transform (FFT) and the decomposition of a given power spectrum into Zernike polynomials
provide
accurate results, though in distinct spatial domains.
While Fourier phase screens resolve the finer structure of phase fluctuations, Zernike screens accurately render the long-range correlations of turbulence
which govern
significant effects in dynamic media \cite{Bachmann23}.

Recently, 
various
combinations of Fourier and Zernike phase screens have been proposed \cite{Zhu15, Bachmann23, Wijerathna23} for atmospheric turbulence.
In this article, we 
review existing methods to generate phase screens and
put forward
\emph{hybrid phase screens} based on the Zernike correction of a Fourier screen for arbitrary random 
continua 
with a given
power spectrum
of refractive index fluctuations.
We show explicitly that, both in terms of the phase structure function as well as of a Zernike expansion, hybrid phase screens yield more accurate representations of the corresponding random medium than other methods while requiring fewer computational resources.

\section{Light propagation through random media}
\label{sec:propagation}
\begin{figure}[h]
    \centering
    \includegraphics[width=.9\columnwidth]{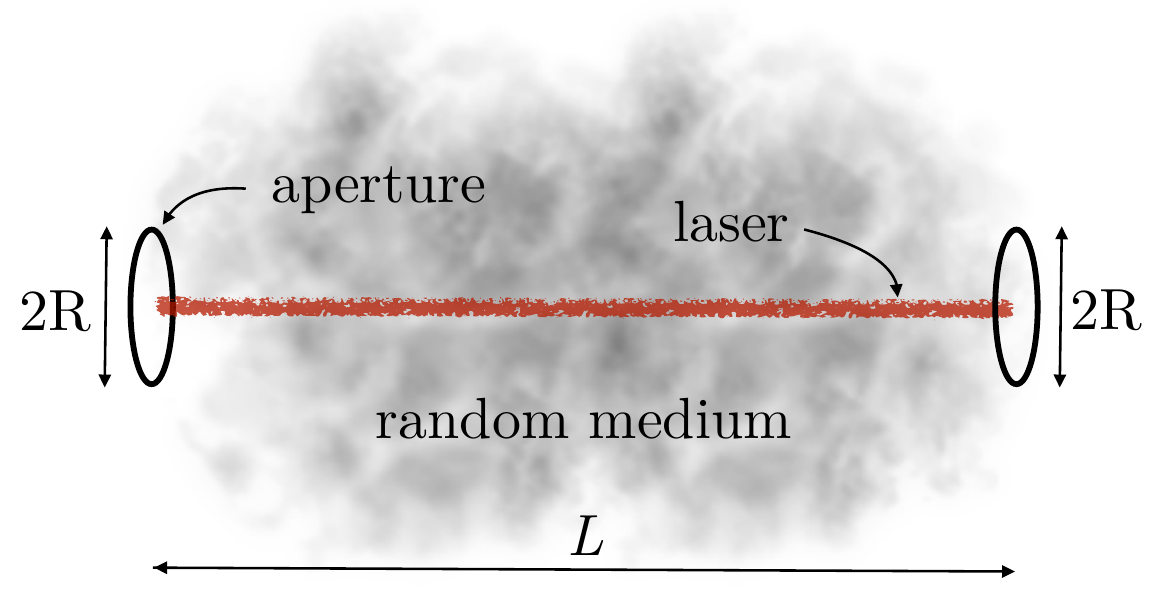}
    \caption{Illustration of the propagation setup. The laser beam propagates from transmitter to receiver, between two coaxial circular apertures of radius $R$. The propagation distance $L$ through the transparent random medium (here drawn partially opaque for illustration) is typically much larger than the aperture sizes, i.e., $L\gg R$.
    }
    \label{fig:prop}
\end{figure}

We consider the propagation of a monochromatic laser beam that is confined by circular apertures of radius $R$ at the transmitter and receiver, as illustrated in Fig.~\ref{fig:prop}. The propagation distance $L$ is assumed to be much larger than the beam's 
width
which is 
limited
by the apertures, i.e., $L\gg R$.
In cases of vacuum propagation, we employ the paraxial approximation of the Helmholtz equation which results in a parabolic equation describing the propagation of light in vacuum \cite{Goodman00}.
If a nonabsorbing, clear of scatterers, random continuum\footnote{E.g., for atmospheric turbulence, the wavelength of the propagated beam may be chosen such that it corresponds to the transparency window of the atmosphere. In particular, infrared wavelengths of $\lambda\approx1550$\,nm 
experience
power losses of less than 1\,\% due to absorption or particle scattering in clear air when propagated over several kilometers \cite{Andrews05}.} is placed between the apertures, the beam's complex amplitude obeys the stochastic parabolic equation. 

\subsection{Stochastic parabolic equation}
Explicitly, the stochastic parabolic equation for a scalar wave $\psi(\bf{r})$ reads \cite{Andrews05}
\begin{equation}
    - 2ik\,\frac{\partial \psi(\bf{r})}{\partial z} =  \Delta_\perp \psi({\bf r}) + 2k^2\,\delta n({\bf r}, t)\, \psi({\bf r}),
    \label{eq:spe}
\end{equation}
where $k$ is the optical wavenumber, $\Delta_\perp$ is the transverse Laplace operator (with respect to the propagation axis $z$), and $\delta n({\bf r}, t)$ is the possibly time-dependent fluctuating part of the refractive index.

\subsection{Split-step method}
\label{sec:split-step}
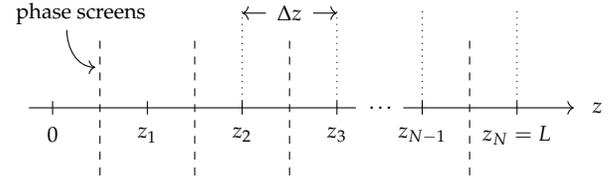
\begin{figure}[h]
    \centering
\begin{tikzpicture}
			[x={(.7, 0)}, y={(0,1)}, z={(0,0)}, scale=.9, every node/.style={scale=.9}]
			
			\draw [-] (-.5,0,0) -- (6.4,0,0);
			\draw (6.95,0,0) node[] {\dots};
			\draw [->] (7.4,0,0) -- (11,0,0);
			
			\draw [-] (0,-.1,0) -- (0,0.1,0);
			\draw [-] (2,-.1,0) -- (2,0.1,0);
			\draw [-] (4,-.1,0) -- (4,0.1,0);
			\draw [-] (6,-.1,0) -- (6,0.1,0);
			\draw [-] (7.8,-.1,0) -- (7.8,0.1,0);
			\draw [-] (9.8,-.1,0) -- (9.8,0.1,0);
			
			\draw [-,dashed] (1,-1,0) -- (1,1,0);
			\draw [-,dashed] (3,-1,0) -- (3,1,0);
			\draw [-,dashed] (5,-1,0) -- (5,1,0);
			\draw [-,dashed] (8.8,-1,0) -- (8.8,1,0);
			
			\draw (0,-.4,0) node[] {$0$};
			\draw (2,-.4,0) node[] {$z_1$};
			\draw (4,-.4,0) node[] {$z_2$};
			\draw (6,-.4,0) node[] {$z_3$};
			\draw (7.8,-.4,0) node[] {$z_{N-1}$};
			\draw (9.8,-.4,0) node[] {$z_{N}=L$};
			\draw (11.5,0,0) node[] {$z$};
			\draw (.6,1.4,0) node[] {phase screens};
			
			\draw (.9, .6) edge[out=180,in=-90,<-] (.3, 1.15);
			
			\draw [-,dotted] (4,.1,0) -- (4,1.5,0);
			\draw [-,dotted] (6,.1,0) -- (6,1.5,0);
			\draw [<-] (4,1.4) --(4.5,1.4);
			\draw [->] (5.5,1.4) --(6,1.4);
			\draw [-,dotted] (7.8,.1,0) -- (7.8,1.5,0);
			\draw [-,dotted] (9.8,.1,0) -- (9.8,1.5,0);
			\draw (5,1.4,0) node[] {$\Delta z$};
		\end{tikzpicture}
    \caption{Schematic of the multiple phase screen model. The propagation path of length $L$ is segmented into diffraction intervals $\Delta z$ each with a centered phase screen.}
    \label{fig:psm}
\end{figure}

In practice, the propagation times
are much shorter than the medium's correlation times,
such that the optical wave experiences a ``frozen'' random realization of turbulence. 
In static turbulence, \eqref{eq:spe} can be solved numerically  by a \emph{split-step method} \cite{Schmidt10,Lukin02,Sorelli19}, which relies on segmenting the propagation path into discrete, medium-induced phase modulations, i.e., \emph{phase screens}. For horizontal links, the screens are separated by equal intervals, as illustrated in Fig.~\ref{fig:psm}; between the screens, the wave undergoes deterministic diffraction.
For example, in atmospheric turbulence, the spacing $\Delta z$ of phase screens has to be chosen sufficiently small, such that for propagation distances smaller than $\Delta z$ scintillation effects are negligible \cite{Schmidt10}. 

\section{Statistical characterization of random media}
\label{sec:power_spectra}
We assume that the 
refractive index fluctuations of the
considered stochastic medium 
are fully
characterized by the 
power spectrum representing its spectral energy distribution. In the following, we
illustrate
this concept
by
the 
famous
\emph{Kolmogorov spectrum}
of
atmospheric turbulence.
Subsequently, we investigate more elaborate atmospheric spectra as well as underwater turbulence.
In fact,
we furnish a recipe to generate phase screens 
for any given power spectrum.

\subsection{Kolmogorov power spectrum}
\begin{figure}
    \centering
    \includegraphics[width=\columnwidth]{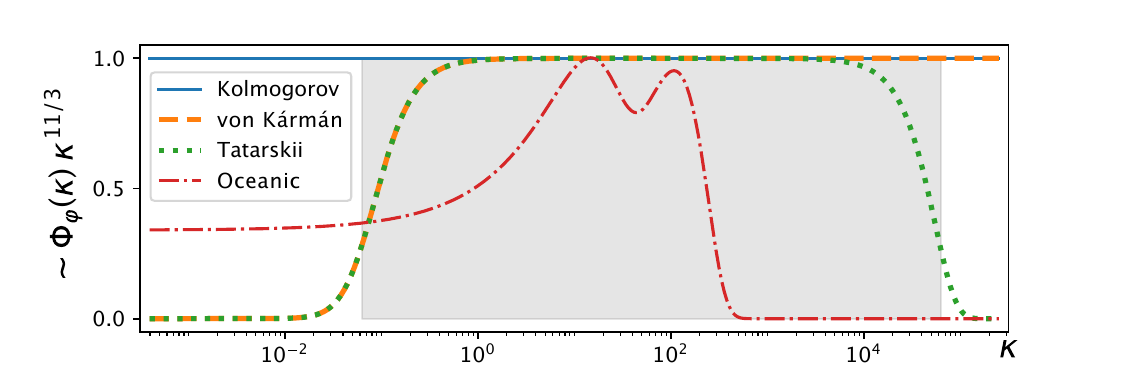}
    \caption{Exemplary plot of different power spectra (see legend). The two-peak structure of the oceanic power spectrum originates from temperature (left peak) and salinity (right peak) fluctuations in sea water. The gray shaded area approximates the inertial range of atmospheric turbulence where length scales are in between the outer and inner scale, that is $l_0 < 2\pi/\kappa < L_0$.
    }
    \label{fig:spectra}
\end{figure}

In the Earth's atmosphere, turbulence stems from small temperature and pressure fluctuations, which amount to the extremely complex dynamics of the turbulent eddies at different spatial scales. Based on an energy cascade hypothesis and dimensional analysis, one of the great achievements of Kolmogorov's theory of turbulence is an expression for the energy of an eddy as a function of its spatial wavenumber \cite{Kolmogorov41a}. Translating Kolmogorov's result into optical turbulence
leads to the three-dimensional power spectrum of the refractive index $n$,
\begin{align}
    \Phi_n^{\text{Kol}}(K) &=\frac{\sqrt{3}\,\Gamma\left(\frac{8}{3}\right)}{8\pi^2}\,C_n^2\, K^{-11/3} \simeq 0.033\, C_n^2\, K^{-11/3}, 
    \label{eq:kol_spec}
\end{align}
where $C_n^2$ is the refractive index structure constant, $K=|\boldsymbol{K}|$ with $\boldsymbol{K}$ being the three-dimensional spatial frequency vector,
and $\Gamma\left(\cdot\right)$ denotes the gamma function \cite{Abramowitz65}.
Equation~(\ref{eq:kol_spec}) is valid within the \emph{inertial range} of turbulence, that is for length scales that are confined by an inner and an outer scale as detailed in the following, cf. Fig.~\ref{fig:spectra} and Secs.~3.\ref{sec:VKS} and 3.\ref{sec:VKTS}.
For our application, it is useful to express \eqref{eq:kol_spec} directly in terms of phase fluctuations of a light beam which propagates through the medium. For a sufficiently short, homogeneous and isotropic turbulent channel of length 
$\Delta z$ 
and for a monochromatic optical wave with wavenumber $k$, 
the power spectrum of phase fluctuations 
$\Phi_\varphi(\kappa)$ is given by \cite{Andrews05}
\begin{equation}
    \Phi_\varphi(\kappa)=2\pi\, k^2\, \Delta z\, \Phi_n(\kappa),
    \label{eq:phiphi}
\end{equation}
with
$\kappa$ 
being
the transverse (with respect to the propagation direction) spatial frequency.
Substitution of \eqref{eq:kol_spec} into \eqref{eq:phiphi} yields the \emph{Kolmogorov phase power spectrum}
\begin{align}
    \Phi_\varphi^{\text{Kol}}(\kappa) &= C_\varphi\, r_0^{-5/3}\, \kappa^{-11/3}\simeq 0.49\, r_0^{-5/3}\, \kappa^{-11/3},
    \label{eq:kol_phase}
\end{align}
with the constant
\begin{equation}
    C_\varphi:=\frac{{2^{2/3} \Gamma^2\left(\frac{11}{6}\right)}}{{\pi^2}} \left[\frac{{24}}{{5}} \Gamma\left(\frac{6}{5}\right)\right]^{5/6} \simeq 0.49,
    \label{eq:cphi}
\end{equation}
and the \emph{Fried parameter}
\begin{align}
    \label{eq:fried}
    r_0:&=(\beta \, k^2\, C_n^2\, \Delta z)^{-3/5},\\\nonumber
    \text{with }\beta&=\frac{\sqrt{3\pi}\Gamma\left(\frac{4}{3}\right)}{2\Gamma\left(\frac{11}{6}\right)\left[\frac{{24}}{{5}} \Gamma\left(\frac{6}{5}\right)\right]^{-5/6}} \simeq 0.42,
\end{align}
which gives the characteristic transverse correlation length of turbulence-induced phase distortions of a plane wave.
The phase power spectrum as given for Kolmogorov turbulence in \eqref{eq:kol_phase} will be the starting point for the generation of phase screens as described in the next section.
The accuracy of given phase screens can be validated by the \emph{phase structure function}
\begin{equation}
    D_\varphi(\Delta r) := \Big\langle\, \overline{\left[ \varphi(r + \Delta r) - \varphi(r)\right]^2}\, \Big\rangle,
    \label{eq:structureDef}
\end{equation}
which describes the average phase variance between two points in the transverse plane separated by a distance of $\Delta r$. The bracket $\langle\cdot\rangle$ denotes \emph{ensemble averaging}, i.e., the average over a large number of specific disorder realizations of a given random medium, and the overline represents the average over all points in the transverse plane.
The structure function of the here considered isotropic media
depends only on the transverse separation $\Delta r$ \cite{Andrews05}.
On the one hand, its instructive definition, given in \eqref{eq:structureDef}, allows one to directly compute a particular structure function for a given ensemble of phase screens (where each screen corresponds to a single realization of disorder), either by means of a Fourier analysis \cite{Schmidt10, Wijerathna23} or by straightforward sampling.
On the other hand, the theoretically expected shape of the structure function may be easily computed from a given phase power spectrum $\Phi_\varphi(\kappa)$ \cite{Ishimaru78,Schmidt10}, i.e.,
\begin{equation}
    D_\varphi(\Delta r) = 4 \pi \int_0^\infty \kappa\, \Phi_\varphi(\kappa)\left[1 - J_0(\kappa\, \Delta r)\right] d\kappa,
    \label{eq:spec2struc}
\end{equation}
where $J_0(\cdot)$ denotes the zeroth order Bessel function of the first kind \cite{Abramowitz65}.
In the case of Kolmogorov turbulence, the integration in \eqref{eq:spec2struc} yields the result
\begin{equation}
    D_\varphi^{\text{Kol}}(\Delta r) = 2\left[\frac{24}{4}\Gamma\left(\frac{6}{5}\right)\right]^{5/6} \left(\frac{\Delta r}{r_0}\right)^{5/3} \simeq 6.88 \left(\frac{\Delta r}{r_0}\right)^{5/3}.
    \label{eq:kol_struc}
\end{equation}

\subsection{Von Kármán spectrum} 
\label{sec:VKS}
As seen from \eqref{eq:kol_phase} as well as from Fig.~\ref{fig:spectra}, 
when employing the Kolmogorov power spectrum (solid blue curve) outside the inertial range (gray shaded area), we will obtain
contributions from arbitrarily small and large spatial frequencies $\kappa$. The inverse, $1/\kappa$, of these frequencies is proportional to the size of turbulent eddies. In reality, one observes, due to inertia, a maximal eddy size corresponding to a finite \emph{outer scale} $L_0$, with values ranging from centimeters to several hundreds of meters (typically bounded by the height over ground) \cite{Andrews05}. This effect is accounted for by the \emph{von Kármán spectrum}
\begin{equation}
    \Phi_\varphi^{\text{Kar}}(\kappa)= C_\varphi\, r_0^{-5/3}\, (\kappa^2 + \kappa_0^2)^{-11/6},
    \label{eq:kar_spec}
\end{equation}
where  $\kappa_0=2\pi/L_0$.
This modification of the Kolmogorov spectrum suppresses low spatial frequencies, which is also seen in Fig.~\ref{fig:spectra} (dashed orange curve).
By plugging \eqref{eq:kar_spec} into \eqref{eq:spec2struc}, we obtain a closed-form expression of the corresponding structure function
\begin{align}
    D_\varphi^{\text{Kar}}(\Delta r)&=2 \frac{\Gamma\left(11/6\right)}{2^{5/6}\pi^{8/3}} \left[\frac{24}{4}\Gamma\left(\frac{6}{5}\right)\right]^{5/6} \left[\frac{2\pi}{r_0 \,\kappa_0}\right]^{5/3}\nonumber\\
    &\times \left[\frac{\Gamma\left(5/6\right)}{2^{1/6}}-\left(\Delta r \,\kappa_0\right)^{5/6}\,K_{5/6}\left(\Delta r\, \kappa_0\right)\right],
    \label{eq:kar_struct}
\end{align}
where $K_{5/6}(\cdot)$ denotes the $5/6$-th order modified Bessel function of the second kind \cite{Abramowitz65}.

\subsection{Tatarskii spectrum}
\label{sec:VKTS}
Finally, also the unbound high spatial frequency contributions of the Kolmogorov or von Kármán spectrum are suppressed in turbulence: The minimal eddy size is limited by viscous energy dissipation leading to a finite \emph{inner scale} $l_0$ in the range of millimeters \cite{Andrews05}. This length scale was first introduced by Tatarskii \cite{Tatarskii16} and may be also considered in presence of infinite outer scales. Here, we directly consider the case of finite inner and outer scales, yielding the following expression:
\begin{equation}
    \Phi_\varphi^{\text{Tat}}(\kappa)=  C_\varphi\, r_0^{-5/3}\, (\kappa^2 + \kappa_0^2)^{-11/6} \exp\left(-\frac{\kappa^2}{\kappa_m^2}\right),
    \label{eq:tat_spec}
\end{equation}
where
$\kappa_m=\left[\sqrt{3}\Gamma\left(8/3\right)/(8\pi)\right]^{-3/4}/ l_0 \simeq 5.48/l_0$.
For brevity, we refer to \eqref{eq:tat_spec} as the \emph{Tatarskii spectrum} which is also known as the von Kármán-Tatarskii spectrum.
The dotted green curve in Fig.~\ref{fig:spectra} illustrates the exponential damping of high spatial frequencies for \eqref{eq:tat_spec}.
On the downside, the integral in \eqref{eq:spec2struc} cannot be solved analytically for the Tatarskii spectrum; however, the exact shape of the corresponding structure function can be obtained numerically.

\subsection{Non-Kolmogorov spectra}
The Kolmogorov power spectrum, as given in \eqref{eq:kol_phase},
can be generalized to other exponents $0<\alpha<2$ \cite{Stribling95, Rao00, Toselli08}:
\begin{equation}
    \Phi_\varphi^\alpha(\kappa) = C_\varphi^\alpha r_0^{-\alpha} \kappa^{-\alpha-2},
    \label{eq:kol_spec_alpha}
\end{equation}
with the modified Fried parameter, c.f.\eqref{eq:fried},
\begin{equation}
    r_0\propto(k^2\, C_n^2\, \Delta z)^{-1/\alpha}.
\end{equation}
By means of \eqref{eq:spec2struc}, the associated structure function reads
\begin{equation}
    D_\varphi^\alpha(\Delta r)=-\frac{2^{1-\alpha}C_\varphi^\alpha \pi\Gamma\left(-\alpha/2\right)}{\Gamma\left(1+\alpha/2\right)}\left(\frac{\Delta r}{r_0}\right)^\alpha,
    \label{eq:nKstruc}
\end{equation}
which 
reduces to \eqref{eq:kol_struc} for the exponent $\alpha=5/3$ corresponding to the Kolmogorov spectrum. In addition, some authors consider 
the quadratic approximation \cite{Stribling95}, i.e., $\alpha=2$.
Since the structure function diverges in this case, as seen from \eqref{eq:nKstruc} [$\Gamma(\cdot)$ diverges for negative integers], we treat the quadratic approximation as the limit $\alpha \to 2$.
In the same manner, also the von Kármán  and Tatarskii spectrum, i.e., Eqs.~(\ref{eq:kar_spec}) and (\ref{eq:tat_spec}), can be generalized to other exponents \cite{Rao00}.

\subsection{Oceanic power spectrum}
The underlying mechanism of underwater turbulence is similar to atmospheric turbulence, but the interplay of temperature and salinity fluctuations in sea water results in a more complicated power spectrum. The combination of the corresponding two scalar spectra yields a high-power polynomial \cite{Ruddick79}. However, for light propagation its linearized form \cite{Nikishov00} has proven to be valid
\cite{Korotkova19}, that is
\begin{align}
    \Phi_\varphi^{\text{oce}}(\kappa)&=\frac{C_0 \alpha_T^2 \chi_n \varepsilon^{-1/3}}{4\pi}\, 
    \kappa^{-11/3} \left[1+C_1 \left(\kappa\,l_0\right)^{2/3}\right] \nonumber\\
    &\times \left[\omega^2 e^{-A_T\delta(\kappa)} + e^{-A_S\delta(\kappa)}- 2\,\omega\, e^{-A_{TS}\delta(\kappa)}\right],
    \label{eq:oce_spec}
\end{align}
where $\delta(\kappa):=\frac{3}{2}C_1^2(\kappa\,l_0)^{4/3}+C_1^3(\kappa\, l_0)^2$ with the inner scale $l_0$ and further constants\footnote{Note that the constant prefactors in \eqref{eq:oce_spec} may also be absorbed in a quantity $r_0$, which can be defined analogously to the previously introduced  Fried parameter in atmospheric turbulence [see \eqref{eq:fried}], i.e., such that it rescales the transverse correlation length of turbulence-induced phase distortions.} summarized in Tab.~\ref{tab:oce_const} in Sec.~\ref{sec:appendix}, Appendix \ref{subsec:oce_paras}. The \emph{mixing parameter} $\omega$ governs the relative contributions of the temperature and salinity spectra: For $\omega \to -\infty$, \eqref{eq:oce_spec} reduces to a pure temperature spectrum while for $\omega \to 0$, we obtain only contributions from the salinity; the two-peak structure in Fig.~\ref{fig:spectra} corresponds to $\omega=-0.8$.
Due to the rather complex form of the spectrum, it is not possible to obtain an analytical expression for its phase structure function $D_\varphi(\Delta r)$, recall \eqref{eq:spec2struc}, which hence has to be evaluated numerically.

\section{Phase screen generation methods}
\label{sec:psm}

In this section, we investigate and benchmark different methods for the generation of phase screens, which are constitutive to the split-step method.
For all spectra that were introduced in the previous section, we generate phase screens for different values of their key parameters.
We set out with the well-known Zernike and Fourier phase screens, and subsequently combine them to construct hybrid phase screens.

\subsection{Zernike phase screens}
\label{subsec:zk}

The use of \emph{Zernike polynomials} for the description of optical wavefronts is well known and their application to atmospheric turbulence traces back to first results by Noll \cite{Noll76} and Roddier \cite{Roddier90}.
These polynomials allow for an accurate representation of
a given energy spectrum, especially for the low-order aberrations, such as tip and tilt.
Furthermore, an adaptive optics correction scheme may be straightforwardly simulated by leaving out the Zernike orders corresponding to the implemented corrections.

\subsubsection{Zernike polynomials}

\begin{figure}[h]
    \centering
    \includegraphics[width=.8\columnwidth]{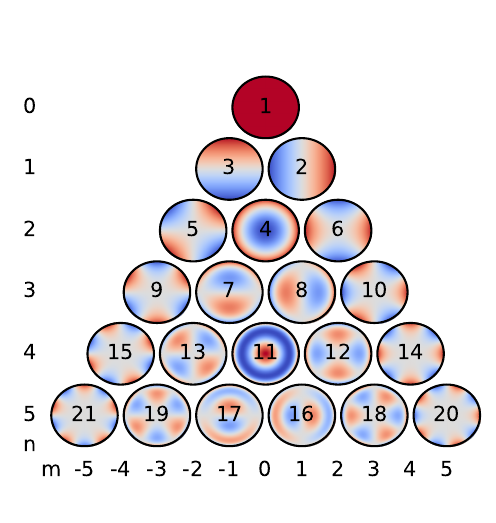}
    \caption{Plot of the first 21 Zernike polynomials $Z_j(\cdot)$ corresponding to radial orders $n\le 5$. The mode ordering index $j$ (black centered numbers) is obtained from \eqref{eq:j}, according to Noll's index convention \cite{Noll76}.
    }
    \label{fig:zernike_pyramid}
\end{figure}

The Zernike polynomials $Z_j(\cdot)$ provide a sequence of orthogonal polynomials within the unit circle
and they are typically expressed as a product of radial and angular functions, that is, with Noll's ordering index $j=j(n,m)$ defined in \eqref{eq:j} below \cite{Noll76}:
\begin{align}
\label{eq:zernike}
     &\left. \begin{array}{lll}
        &Z_{\text{even }j}(r,\theta)&=\sqrt{n+1}\,R_n^m(r)\,\sqrt{2}\cos(m\theta) \\
        &Z_{\text{odd }j}(r,\theta)&=\sqrt{n+1}\,R_n^m(r)\,\sqrt{2}\sin(m\theta)
    \end{array}\,\right\} m\neq 0,\nonumber\\
    &\left.\begin{array}{lll} &Z_{j}(r,\theta)_{\hphantom{\text{even }}} &= \sqrt{n+1}\,R_n^0(r), \\ \end{array}\right. \qquad\qquad\quad\,\, m= 0,
\end{align}
where
\begin{equation}
    R_n^m(r)=\hspace{-8pt}\sum_{s=0}^{(n-|m|)/2}\hspace{-8pt}\frac{(-1)^s\,(n-s)!}{s!\,\left[(n+m)/2 -s\right]!\,\left[(n-m)/2-s\right]!}\,r^{n-2s}.
\end{equation}
An overview of the first 21 Zernike polynomials is provided in Fig.~\ref{fig:zernike_pyramid}.
The radial index $n\in\mathbb{N}_0$ and the azimuthal index $m\in\mathbb{Z}$ satisfy the relations $|m|\leq n$, and $n - |m|$ is even.
Noll's \emph{mode ordering index} $j=j(n,m)$ is given by
\begin{equation}
    j := \frac{n(n+1)}{2}+|m|+\left\{\begin{array}{ll}
0, & m>0 \land (n \text{\,mod\,} 4) \in \{0,1\},\\
0,  & m<0 \land (n \text{\,mod\,} 4) \in \{2,3\},\\
1,  & m \ge 0 \land (n \text{\,mod\,} 4) \in \{2,3\},\\
1, & m \le 0 \land (n \text{\,mod\,} 4) \in \{0,1\},
\end{array}\right.
\label{eq:j}
\end{equation}
and illustrated in Fig.~\ref{fig:zernike_pyramid}.
Such mode ordering in particular allows to formulate the orthogonality relation
\begin{equation}
    \int_0^\infty r\, dr \int_0^{2\pi} d\theta\,  W(r)\, Z_j(r,\theta)\, Z_{j'}(r,\theta) = \delta_{jj'},
\end{equation}
where $W(\cdot)$ is the aperture weighting function
\begin{equation}
    W(r) = \begin{cases} 1/\pi, &r\leq 1,\\
    0,  &r>1,
    \end{cases}
\end{equation}
and $\delta_{jj'}$ is the Kronecker delta.
Thus, we may expand an arbitrary (here real) wavefront $\varphi(r,\theta)$ over a circular \emph{aperture with radius} $R$ in terms of Zernike polynomials, i.e.,
\begin{equation}
    \varphi(R\rho, \theta)= \sum_{j=1}^\infty a_j\, Z_j(\rho, \theta),
    \label{eq:zernike_exp}
\end{equation}
with $\rho:=r/R$ and the \emph{Zernike expansion coefficients}
\begin{equation}
    a_j=\int_0^\infty \rho\, d\rho \int_0^{2\pi} d\theta\, W(\rho) \,\varphi(R\rho, \theta)\, Z_j(\rho, \theta).
    \label{eq:aj}
\end{equation}
Finally, we introduce the Fourier transform $Q_j(\boldsymbol{\kappa}=\kappa,\phi)$ of $Z_j(\boldsymbol{\rho}=\rho,\theta)$ which satisfies
\begin{equation}
     W(\rho)\, Z_j(\boldsymbol{\rho}) = \int d^2 \boldsymbol{\kappa}\, Q_j(\boldsymbol{\kappa})\, e^{-2\pi i\, \boldsymbol{\kappa}\cdot\boldsymbol{\rho}},
    \label{eq:filterFT}
\end{equation}
where the integration is performed over the entire frequency space. By \eqref{eq:zernike}, one finds
\begin{align}
\label{eq:qzernike}
     \left. \begin{array}{ll}
        Q_{\text{even }j}&= \sqrt{2}(-1)^{(n-m)/2}i^m\cos (m \phi) \\
        Q_{\text{odd }j}&= \sqrt{2}(-1)^{(n-m)/2}i^m\sin (m \phi) \\
        Q_{j}^{(m=0)}&= (-1)^{n/2}
    \end{array}\hspace{-3pt}\right\} \times \frac{\sqrt{n+1}}{\pi \kappa}J_{n+1}(2\pi \kappa),
\end{align}
where on the left hand side the arguments $\kappa, \phi$ of $Q_j$ were suppressed for brevity.

\subsubsection{Covariance of Zernike expansion coefficients}

The \emph{Wiener-Khinchin theorem} \cite{Wiener30, Khintchine34, Noll76, Wang:78, Roddier90} relates the covariance of the Zernike expansion coefficients $a_j$ of a wavefront distorted by a given random medium, see \eqref{eq:aj}, to the phase power spectrum $\Phi_\varphi(\kappa)$ of such a medium.
Provided that the Zernike coefficients $a_j$ are zero-mean
random variables, their covariance is given by
\begin{align}
    \langle a_j a_{j'}^* \rangle = &\int d^2\boldsymbol{\rho} \int d^2\boldsymbol{\rho}'\, W(\rho)\, Z_j(\rho, \theta)\,\nonumber\\ \times&\langle \varphi(R\rho)\, \varphi(R\rho') \rangle\, W(\rho')\, Z_{j'}(\rho',\theta'),
\label{eq:covariance}
\end{align}
where the brackets denote ensemble averaging [analogous to that in \eqref{eq:structureDef}] and $\langle \varphi(R\rho)\, \varphi(R\rho') \rangle$ yields the phase covariance (where we suppressed its angular arguments $\theta, \theta'$ in the assumption of isotropy).
Next, we rewrite \eqref{eq:covariance} in Fourier space, 
with the help of \eqref{eq:qzernike}:
\begin{align}
    \langle a_j a_{j'}^* \rangle = & \int d^2\boldsymbol{\kappa} \int d^2\boldsymbol{\kappa}' \, Q_j(\boldsymbol{\kappa})\, Q_{j'}^*(\boldsymbol{\kappa}')\\\nonumber
    \times& \int d^2\boldsymbol{\rho} \int d^2\boldsymbol{\rho}'\,\langle \varphi(R\rho)\, \varphi(R\rho') \rangle\, e^{-2\pi i\, \boldsymbol{\kappa}\cdot\boldsymbol{\rho}}\, e^{+2\pi i\, \boldsymbol{\kappa}'\cdot\boldsymbol{\rho}'}.
\end{align}
The second line is the spectral decomposition of the phase covariance which yields, by means of the Wiener-Khinchin theorem, the phase power spectrum $\Phi_\varphi\left(\kappa\right)$. Thus we have
\begin{align}
    \langle a_j a_{j'}^* \rangle = & \int d^2\boldsymbol{\kappa} \int d^2\boldsymbol{\kappa}' \, Q_j(\boldsymbol{\kappa})\,Q_{j'}^*(\boldsymbol{\kappa}') \, \Phi_\varphi\left(\frac{\kappa}{R}, \phi; \frac{\kappa'}{R}, \phi'\right),
    \label{eq:ajaj_old}
\end{align}
where $\Phi_\varphi\left(\kappa/R, \phi; \kappa'/R, \phi'\right)$ represents a general power spectrum that may depend on $\boldsymbol{\kappa}=(\kappa, \phi)$ and $\boldsymbol{\kappa}'=(\kappa', \phi')$.
For the considered isotropic media, described by locally stationary random processes, 
the induced phase covariance
depends only on the transverse 
distance $\Delta \rho := \rho' - \rho$, and the power spectrum takes the form \cite{Andrews05}
\begin{equation}
    \Phi_\varphi\left(\boldsymbol{\kappa}; \boldsymbol{\kappa}'\right) = \delta^{(2)}(\boldsymbol{\kappa}-\boldsymbol{\kappa}')\, \Phi_\varphi(\kappa),
    \label{eq:stationary}
\end{equation}
with the two-dimensional Dirac delta distribution $\delta^{(2)}(\boldsymbol{\kappa} - \boldsymbol{\kappa}')$,
 which reads, in polar coordinates (for $\kappa, \kappa' \neq 0$)
\begin{equation}
   \delta^{(2)}(\boldsymbol{\kappa} - \boldsymbol{\kappa}') = \frac{1}{\kappa}\,\delta(\kappa - \kappa')\, \delta(\phi - \phi').
    \label{eq:delta2}
\end{equation}
Finally, substituting Eqs.~(\ref{eq:stationary}) and (\ref{eq:delta2}), together with the Fourier-transformed Zernike polynomials, \eqref{eq:qzernike}, into the Zernike covariance given by \eqref{eq:ajaj_old}, we obtain
\begin{align}
    \langle a_j a_{j'}^* \rangle &= 8\pi \sqrt{(n+1)(n'+1)}\,(-1)^{(n+n'-m-m')/2}\nonumber\\ &\times\int_0^\infty d\kappa\, \kappa\, \Phi_\varphi(\kappa) \frac{J_{n+1}\left(R \kappa\right)}{R \kappa}\, \frac{J_{n'+1}\left(R \kappa\right)}{R \kappa}\, \tilde{\delta}_{jj'},
    \label{eq:ajaj}
\end{align}
where $\tilde{\delta}_{jj'}$ denotes the logical Kronecker symbol:
\begin{equation}
    \tilde{\delta}_{jj'}=\begin{cases}
        1, (m=m') \land \left[\left(\frac{j - j'}{2} \text{ even}\right) \lor (m=0)\right],\\
        0, \text{ else.}
    \end{cases}
    \label{eq:tdelta}
\end{equation}
For the Kolmogorov and the von Kármán spectrum, the integral in \eqref{eq:ajaj} can be solved analytically, and the corresponding results are listed in Sec.~\ref{sec:appendix}, Appendix \ref{subsec:aj_ana}.

\begin{figure}
    \centering
    \includegraphics[width=.8\columnwidth]{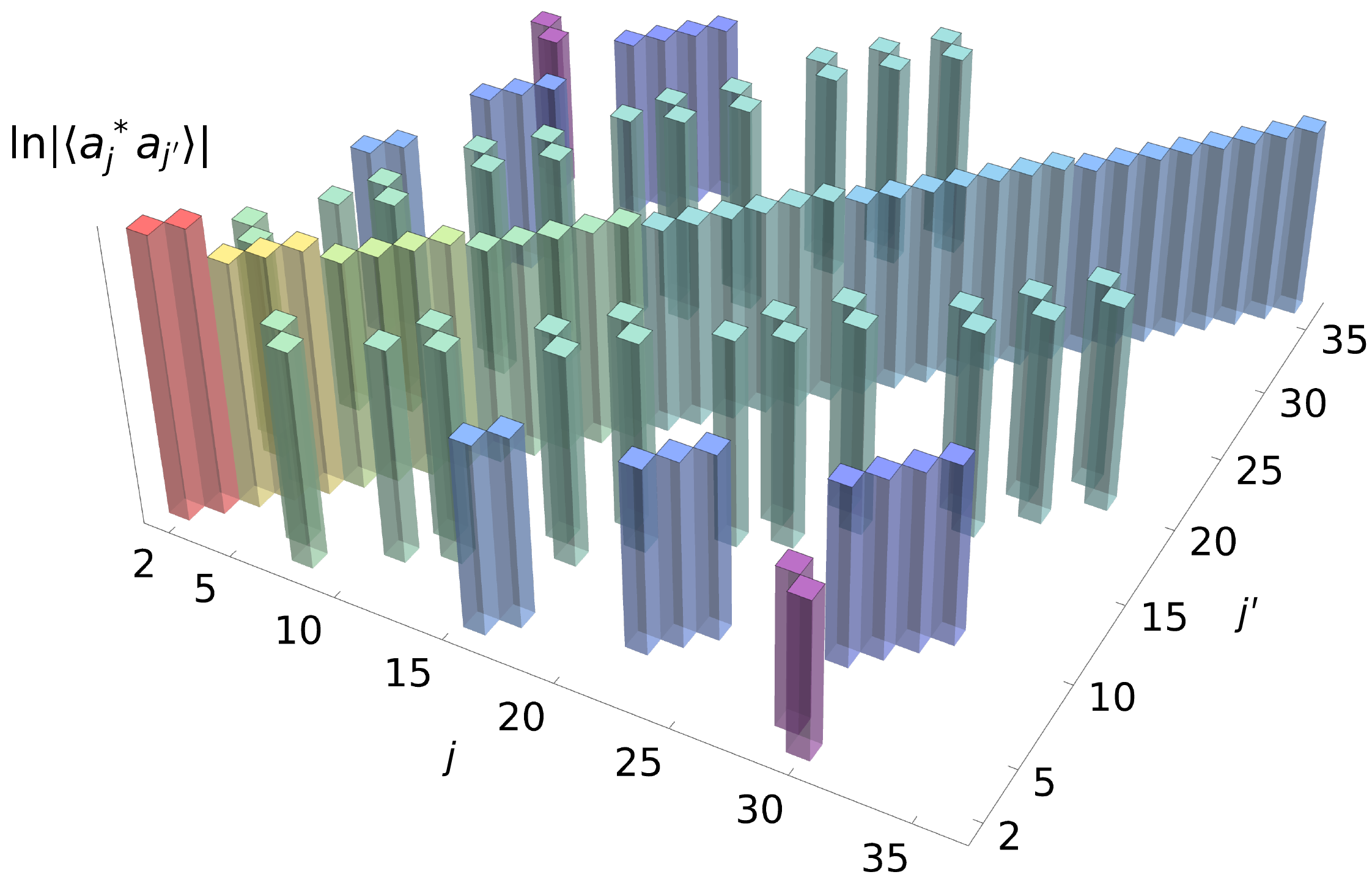}
    \caption{Bar plot of the Zernike coefficient covariance $\langle a_j a_{j'}^*\rangle$ for oceanic turbulence as given in \eqref{eq:ajaj}. Note that, due to the logarithmic scale, the absolute value is plotted.
    }
    \label{fig:ajaj}
\end{figure}

\subsubsection{Generation of Zernike phase screens}
\label{sssec:generationZ}

The general idea of a Zernike phase screen is to generate a random sequence of Zernike coefficents whose covariances obey the statistics prescribed by a given power spectrum.
However, since the Zernike polynomials given in \eqref{eq:zernike} are not statistically independent, their expansion coefficients $a_j$ [see \eqref{eq:aj}] exhibit in general non-vanishing covariances, i.e., $\langle a_j a_{j'}^*\rangle$ given in \eqref{eq:ajaj} does not vanish for all $j\neq j'$ as illustrated in Fig.~\ref{fig:ajaj}.
Thus we perform a singular value decomposition (SVD) yielding the orthonormal \emph{Karhunen-Loève functions}
of the covariance matrix $\langle a_j a_{j'}^*\rangle$
in order to obtain independent random variables \cite{Roddier90, roggemann2018imaging}.
Unfortunately, there exists no analytical expression for these functions.

For convenience, let us collect the Zernike expansion coefficients
in the form of a column vector, that is
\begin{equation}
    \boldsymbol{A}:=(a_2, a_3, \dots, a_J)^T,
\end{equation}
where $J$ is the \emph{highest considered order of Zernike polynomials}
and $T$ denotes transposition\footnote{We do not list $a_1$, since this ``piston'' mode corresponds to a global phase shift
.}.
Evidently, the consideration of only a finite number of Zernike polynomials introduces a systematic error, since the expansion coefficients decay only algebraically with the mode order, cf. Fig.~\ref{fig:ajaj}.
This deficiency will be overcome later in our definition of hybrid phase screens.
By construction, the covariance matrix $\langle \boldsymbol{A}\boldsymbol{A}^\dagger \rangle$ [where the bracket denotes, as previously seen in Eqs.~(\ref{eq:structureDef}) and  (\ref{eq:covariance}), ensemble averaging] is Hermitian, allowing its
SVD as follows:
\begin{equation}
    \langle \boldsymbol{A}\boldsymbol{A}^\dagger \rangle = U^\dagger S U,
    \label{eq:svd}
\end{equation}
where
$U$ is a unitary matrix, $S$ is diagonal, 
and $\boldsymbol{A}\boldsymbol{A}^\dagger$ describes the matrix multiplication between the column vector $\boldsymbol{A}$ and its transposed complex conjugate $\boldsymbol{A}^\dagger$.
Applying the matrix $U$ to the original Zernike coefficients $\boldsymbol{A}$ then yields the statistically independent \emph{Karhunen-Loève coefficients} given by
\begin{equation}
    \boldsymbol{B}=U \boldsymbol{A},
    \label{eq:bs}
\end{equation}
and their covariance matrix is---as desired---diagonal,
\begin{equation}
    \langle \boldsymbol{B}\boldsymbol{B}^\dagger \rangle = \langle U\boldsymbol{A} \boldsymbol{A}^\dagger U^\dagger\rangle = U \langle \boldsymbol{A} \boldsymbol{A}^\dagger \rangle U^\dagger = S,
    \label{eq:bcov}
\end{equation}
where we have used \eqref{eq:svd} in the last step.

Consequently, to obtain the Zernike expansion coefficients corresponding to a given power spectrum, we start by generating a vector $\boldsymbol{B}$ of $J-1$ normally distributed random variables with zero mean and variances given by the diagonal matrix $S$ corresponding to the Karhunen-Loève coefficients \cite{Segel19}.
These are then easily transformed into the actual Zernike coefficients by inverting \eqref{eq:bs}, that is
\begin{equation}
    \boldsymbol{A} = U^\dagger \boldsymbol{B},
\end{equation}
and the final phase screen is obtained using \eqref{eq:zernike_exp}.

\begin{figure}
    \centering
    \includegraphics[width=\columnwidth]{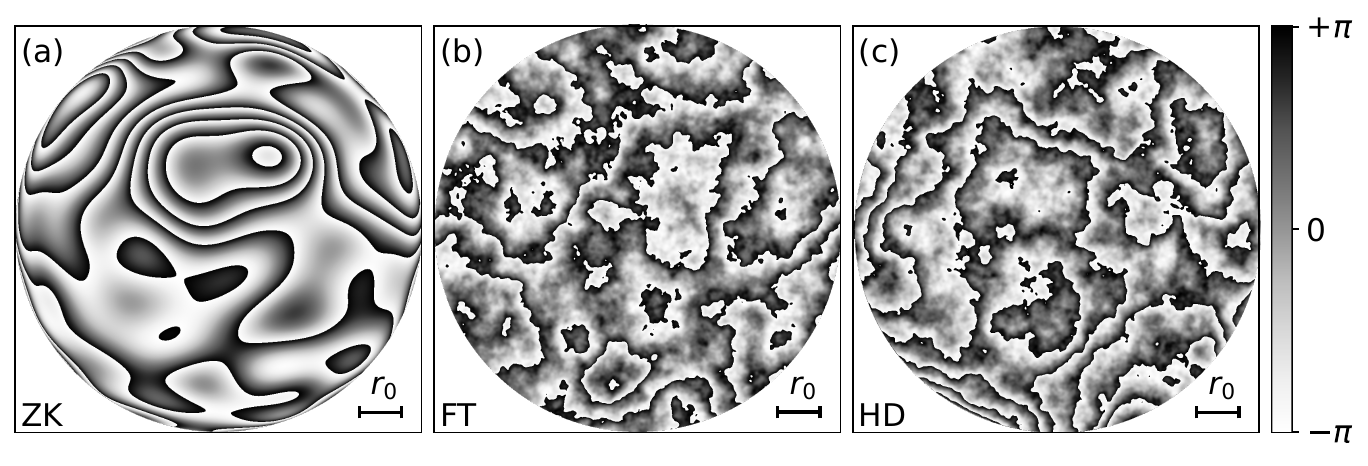}
    \caption{Sample phase screens for oceanic turbulence.
    (a) Zernike phase screen with the expansion in \eqref{eq:zernike_exp} ceasing at $J=120$.
    (b) Fourier phase screen with $10$ subharmonics.
    (c) Hybrid phase screen with $J=21$ Zernike expansion coefficients in \eqref{eq:zernike_exp}.
    }
    \label{fig:phasescreens}
\end{figure}

In practice, depending on the form of the considered spectrum, it is reasonable to factor out, e.g., the Fried parameter $r_0$, or, if possible, the aperture radius $R$, such that the Zernike expansion coefficients may be precomputed and saved for different turbulent channels\footnote{The aperture radius $R$ may be factored out, e.g., for the Kolmogorov spectrum.}.
Furthermore, $J$, i.e., the highest considered order of Zernike polynomials, should be chosen such that full radial orders of Zernike polynomials are considered corresponding to complete rows in Fig.~\ref{fig:zernike_pyramid}.
A sample Zernike (ZK) phase screen is plotted in Fig.~\ref{fig:phasescreens}(a), for $J=120$, that is up to the radial order $n=14$.

\subsubsection{Statistical validation of Zernike phase screens}
\label{sssec:procedure}

\begin{figure}[h]
    \centering
    \includegraphics[width=\columnwidth]{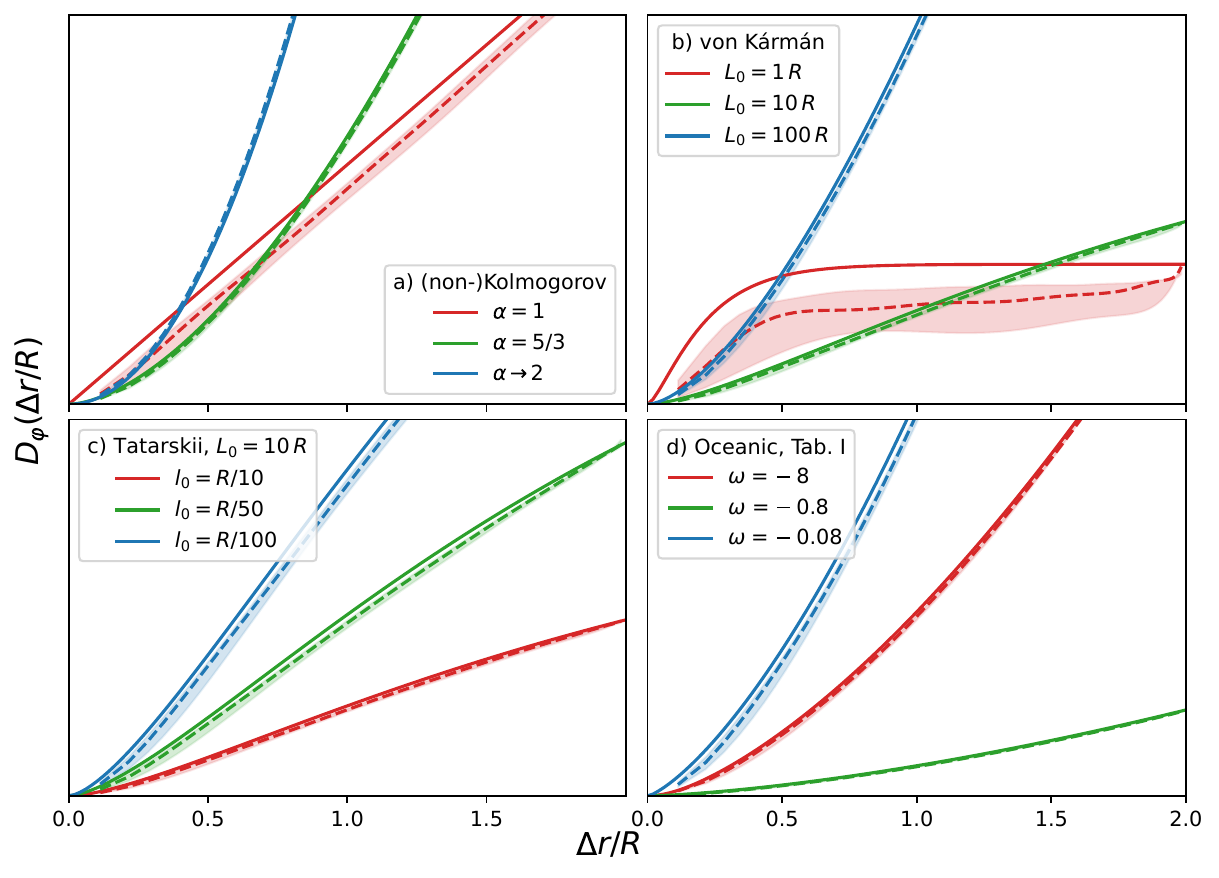}
    \caption{Rescaled structure functions $D_\varphi(\Delta r/R)$, see \eqref{eq:structureDef}, of Zernike phase screens with upper limit $J=21,55,120$ in the expansion \eqref{eq:zernike_exp}, for the (non-)Kolmogorov, von Kármán, Tatarskii and oceanic spectrum, and for three different choices of the relevant spectral parameters (see legend), obtained after ensemble averaging over
    1000 realizations. The theoretically expected structure functions from \eqref{eq:spec2struc} are plotted as solid lines. The dashed lines indicate the performance of Zernike phase screens with $J=55$, and the bounds of the accompanying shaded areas (not resolved in all cases) correspond to $J=21$ and $J=120$, respectively, where the bound closer to the theoretical curve corresponds in all cases to the larger number of considered Zernike expansion coefficients.
    Note that each set of curves, obtained for a given spectral parameter, has been rescaled such that the curves for all three chosen parameters can be shown in a single plot. This allows for qualitative but not quantitative comparison of the structure functions associated with different spectral parameters.
    In practice, such a rescaling may be archived by a variation of the media's Fried parameters. 
    }
    \label{fig:zernike_dphi}
\end{figure}

To quantify the faithfulness with which specific structure functions can be realized by Zernike phase screens, we choose three values for the characteristic parameter of each of the power spectra introduced in Sec.~\ref{sec:power_spectra}, as summarized in Tab.~\ref{tab:spec_paras} in Sec.~\ref{sec:appendix}, Appendix \ref{subsec:sim_paras}.
For each spectrum and each parameter value, we generate
1000 phase screens and their
structure function --
evaluated according to the prescription in \eqref{eq:structureDef}---is
compared to the theoretical structure function that is directly obtained from the underlying power spectrum by \eqref{eq:spec2struc}.
The results are plotted, for three different
values $J$ of the (in our numerical approximation finite) upper limit of the expansion in \eqref{eq:zernike_exp},
in Fig.~\ref{fig:zernike_dphi}.

In almost all cases we find very good agreement with the from \eqref{eq:spec2struc} theoretically expected curves. The deviation is strongest when the outer scale of turbulence is close to the aperture size, i.e., $L_0\approx R$, as seen for von Kármán turbulence in Fig.~\ref{fig:zernike_dphi}(b) (red lines). This behavior is indeed expected, since the choice of such a small outer scale effectively suppresses all long range correlations, which are precisely described by low-order Zernike polynomials. Thus, when considering only a limited number of Zernike expansion coefficients (starting from the lowest order), the absence of high-order, i.e. short-range, Zernike contributions introduces in this case the greatest systematic error. Furthermore, it is notable that even when considering only a very limited number of Zernike polynomials, e.g., $J=21$, we nevertheless find structure functions very close to the theoretical prediction. Their deviation from the theoretical curve is compared to that observed for other phase screen generation methods in Sec.~\ref{sec:psm}.\ref{subsec:comparison}.

\begin{figure}
    \centering
    \includegraphics[width=\columnwidth]{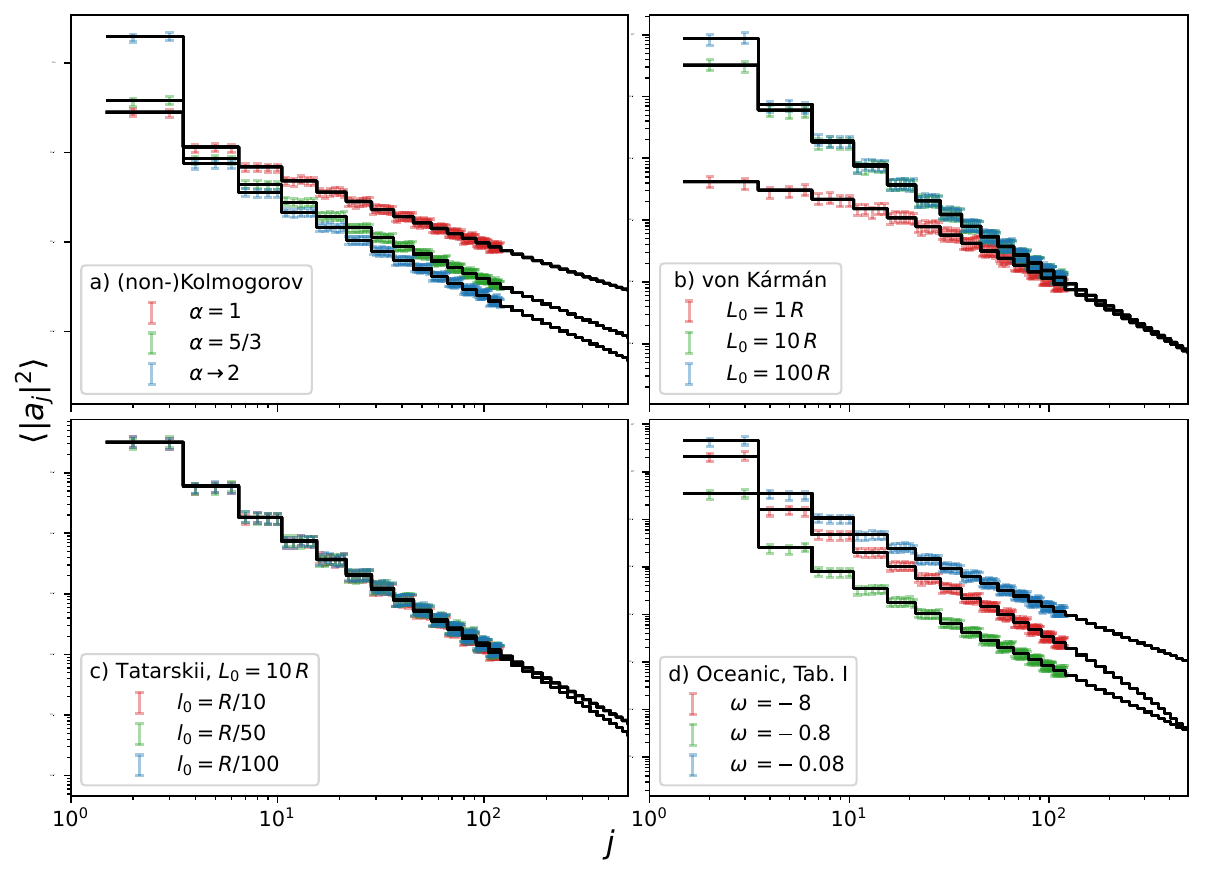}
    \caption{Log-log plot of the variances  $\langle|a_j|^2\rangle$ of the Zernike expansion coefficients (with zero mean) in \eqref{eq:zernike_exp}, for Zernike phase screens with an expansion up to $J=120$ for the four different spectra given in Eqs.~(\ref{eq:kol_phase}), (\ref{eq:kar_spec}), (\ref{eq:tat_spec}) and (\ref{eq:oce_spec}) (see legend). The solid black lines connect the values theoretically expected from \eqref{eq:ajaj}. Error bars correspond to one standard deviation after ensemble averaging
    1000 disorder realizations of the turbulent medium.
    }
    \label{fig:zernike_ajs}
\end{figure}

Finally, we may also decompose the generated phase screens into Zernike polynomials and compare the obtained covariances of expansion coefficients, after ensemble averaging, to their theoretical values as given in \eqref{eq:ajaj}. As shown for the variances $\langle|a_j|^2\rangle$ in Fig.~\ref{fig:zernike_ajs}, for $J=120$, we find---consistently with the results for the structure function---very good agreement with the theoretical values (black lines), up to the highest considered order $J$.
Although, of course, this is expected by construction, it certifies that the ensemble average of 1000 phase screens provides statistically meaningful results.
Furthermore, these plots illustrate graphically the impact of the different chosen spectral parameters. For instance, the exponent $\alpha$ of the generalized Kolmogorov turbulence in (a) governs the overall slope of the monotonically decreasing variance of higher orders. For small outer scales $L_0\approx R$, i.e., here for the von Kármán turbulence (b), we observe that also the variance of the suppressed low-order Zernike contributions (cf. discussion above) is diminished by several orders of magnitude [red points in Fig.~\ref{fig:zernike_ajs}(b)].
In (c), we observe that different inner scales $l_0$ only affect higher Zernike orders representing the fine structure, while the long-ranging lower orders remain---as expected---unaffected. For the oceanic spectrum (d), we obtain distinct variances of Zernike expansion coefficients for different mixing parameters $\omega$, where for dominant temperature contributions variances of higher order coefficients are suppressed (see decay of the theoretical curve for $\omega=-8$). However, this is not resolved in the simulated data since we only included the first $J=120$ Zernike polynomials.

\subsection{Fourier phase screens}
\label{subsec:ft}
An alternative to Zernike phase screens and, likewise, 
widely used method for the numerical generation of phase screens 
relies on the fast Fourier transform (FFT) algorithm \cite{Johansson94}. This method is attractive due to simplicity of its implementation and reasonable accuracy of the resulting screens.

\subsubsection{Spatial phase modulation}
The phase modulation function $\varphi(\boldsymbol{\rho})$ can be obtained in two steps. First, the square root of the power spectrum is multiplied by a spectral response function $g(\boldsymbol{\kappa})$, a procedure called \emph{filtering} in theory of random processes \cite{Gallagher_book}. Second, an inverse Fourier transform from frequency to real space  
results in phase modulation $ \varphi(\boldsymbol{\rho} = \rho_x, \rho_y)$ (here for convenience in Cartesian coordinates) in the transverse plane \cite{Johansson94}
\begin{equation}
    \varphi(\boldsymbol{ \rho}) = \int d^2\boldsymbol{\kappa}\,g(\boldsymbol{\kappa}) \sqrt{\Phi_\varphi(\kappa)} e^{i\boldsymbol{\kappa}\cdot\boldsymbol{\rho}},
    \label{eq:filter}
\end{equation}
where the integration is performed over the entire frequency space with $\boldsymbol{\kappa}=(\kappa_x, \kappa_y)$ and $\kappa=|\boldsymbol{\kappa}|$.
In our present case, $g(\boldsymbol{\kappa})$ indicates Hermitian, zero-mean Gaussian white noise of unit variance---which ensures that a Fourier transform of $\langle \varphi(\boldsymbol{\rho})\varphi(\boldsymbol{\rho}^\prime)\rangle$ yields the power spectrum.
Further details for the numerical implementation of
filtering are given in Sec.~\ref{sec:appendix}, Appendix~\ref{subsec:ft_details}.

Although \eqref{eq:filter} is theoretically exact, the precision of its numerical implementation is limited by discretization and a finite grid size.
From the former, it is well known that the consequent undersampling of low spatial frequencies \cite{Johansson94} suppresses low order aberrations such as tip and tilt which are indeed constituting a major fraction of the turbulent energy spectrum \cite{Roddier90,Noll76}.
Especially for simulations of longer channels this ultimately results in an underestimation of turbulence effects such as beam wandering.
On the other hand, the finite grid size poses problems with resonances due to its boundaries. In order to avoid these artefacts, Fourier screens should always be generated with sufficient padding, as detailed in Sec.~\ref{sec:appendix}, Appendix \ref{subsec:ft_details}.

\subsubsection{Subharmonic Fourier phase screens}

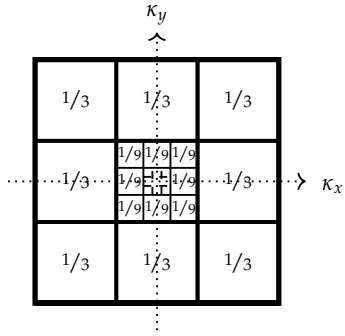
\begin{figure}[h]
		\centering
		\begin{tikzpicture}
			[x={(.2, 0)}, y={(0,.2)}, z={(0,0)}, scale=.6, every node/.style={scale=0.8}]
			
			\draw [line width=0.7mm,] (0,0,0) -- (27,0,0) -- (27,27,0) -- (0,27,0) -- (0,0,0) -- (27, 0,0);
			
			\draw [line width=0.5mm,] (9,0,0) -- (9,27,0);
			\draw [line width=0.5mm] (18,0,0) -- (18,27,0);
			\draw [line width=0.5mm] (0,9,0) -- (27,9,0);
			\draw [line width=0.5mm] (0,18,0) -- (27,18,0);
			
			\draw [line width=0.5mm,] (9,0,0) -- (9,27,0);
			\draw [line width=0.5mm] (18,0,0) -- (18,27,0);
			\draw [line width=0.5mm] (0,9,0) -- (27,9,0);
			\draw [line width=0.5mm] (0,18,0) -- (27,18,0);
			
			\draw [line width=0.25mm,] (12,9,0) -- (12,18,0);
			\draw [line width=0.25mm,] (15,9,0) -- (15,18,0);
			\draw [line width=0.25mm,] (9,12,0) -- (18,12,0);
			\draw [line width=0.25mm,] (9,15,0) -- (18,15,0);
			
			\draw [line width=0.25mm, dashed] (13,12,0) -- (13,15,0);
			\draw [line width=0.25mm, dashed] (14,12,0) -- (14,15,0);
			\draw [line width=0.25mm, dashed] (12,13,0) -- (15,13,0);
			\draw [line width=0.25mm, dashed] (12,14,0) -- (15,14,0);
			
			\draw (4.5,4.5,0) node[] {\Large$\nicefrac{1}{3}$};
			\draw (4.5+9,4.5,0) node[] {\Large$\nicefrac{1}{3}$};
			\draw (4.5+9+9,4.5,0) node[] {\Large$\nicefrac{1}{3}$};
			
			\draw (4.5,4.5+9,0) node[] {\Large$\nicefrac{1}{3}$};
			\draw (4.5+9+9,4.5+9,0) node[] {\Large$\nicefrac{1}{3}$};
			
			\draw (4.5,4.5+18,0) node[] {\Large$\nicefrac{1}{3}$};
			\draw (4.5+9,4.5+18,0) node[] {\Large$\nicefrac{1}{3}$};
			\draw (4.5+9+9,4.5+18,0) node[] {\Large$\nicefrac{1}{3}$};
			
			\draw (10.5,10.5,0) node[] {$\nicefrac{1}{9}$};
			\draw (13.5,10.5,0) node[] {$\nicefrac{1}{9}$};
			\draw (16.5,10.5,0) node[] {$\nicefrac{1}{9}$};
			
			\draw (10.5,13.5,0) node[] {$\nicefrac{1}{9}$};
			\draw (16.5,13.5,0) node[] {$\nicefrac{1}{9}$};
			
			\draw (10.5,16.5,0) node[] {$\nicefrac{1}{9}$};
			\draw (13.5,16.5,0) node[] {$\nicefrac{1}{9}$};
			\draw (16.5,16.5,0) node[] {$\nicefrac{1}{9}$};
			
			\draw (33,27/2-.5,0) node[] {\Large$\kappa_x$};
			\draw (27/2,32,0) node[] {\Large$\kappa_y$};
			
			\draw [->, dotted, line width=0.25mm] (-3,27/2,0) -- (30, 27/2, 0);
			\draw [->, dotted, line width=0.25mm] (27/2,-3,0) -- (27/2, 30, 0);

		\end{tikzpicture}
		\caption{Illustration of the first three subharmonic levels that replace the single sampling at the origin of a phase screen in Fourier space, according to the method proposed in Ref.~\cite{Lane92}. The centered numbers represent the weighting factors $\sqrt{\Delta \kappa_x \Delta \kappa_y}$ due to the smaller sampling area corresponding to a sampling point.}
		\label{fig:subharmonics}
	\end{figure}

Several strategies have been proposed to overcome these issues.
Most prominently, the introduction of \emph{subharmonic levels}, or subharmonics for Fourier phase screens \cite{Herman90,Lane92,Johansson94} strives to compensate the missing low spatial frequencies by additional finer grid layers.
The key idea is to replace the sampling at the origin of the spectral space with nested layers of finer grids which are weighted corresponding to their area.
Figure~\ref{fig:subharmonics} illustrates the subharmonic method that was originally proposed in Ref.~\cite{Lane92}, where the centered sample point at the origin of the spectral grid is replaced by nine sample points. This replacement may be iteratively continued leading to deeper subharmonic levels and more accurate representations of low frequency contributions.
However, we observe, in accordance with other authors \cite{Johansson94, Schmidt10}, that with more than ten subharmonic levels the increased runtime, as well as the accumulation of numerical errors, render such higher order subharmonic screens inconvenient.
Similar results are also found for slightly more refined subharmonic grids, as presented, for example, in Ref.~\cite{Johansson94}.

A sample Fourier transform (FT) phase screen with 10 subharmonics is plotted in Fig.~\ref{fig:phasescreens}(b).
We note that this phase screen shows a visibly more refined structure in comparison to the Zernike phase screen in Fig.~\ref{fig:phasescreens}(a).
This is due to the necessarily finite number of Zernike expansion coefficients which limits the resolution of finer structures.
On the other hand, we observe that the Fourier phase screen in Fig.~\ref{fig:phasescreens}(b) is overall flatter, meaning that we observe in this case approximately four phase revolutions along a path across the aperture, whereas the Zernike phase screen in Fig.~\ref{fig:phasescreens}(a) produces up to nine phase revolutions across the aperture.
This behavior reflects the missing low-order Zernike contributions, corresponding to e.g. tip and tilt, of Fourier phase screens.

\subsubsection{Statistical validation of Fourier phase screens}

As before, we generate for each of the previously introduced spectra
1000 (subharmonic) Fourier phase screens, again with the same three 
values of the characteristic spectral parameters as given in Tab.~\ref{tab:spec_paras} in Sec.~\ref{sec:appendix}, Appendix~\ref{subsec:sim_paras}, and evaluate the phase structure function as well as the Zernike expansion in \eqref{eq:zernike_exp} after ensemble averaging.

\begin{figure}[h]
    \centering
    \includegraphics[width=\columnwidth]{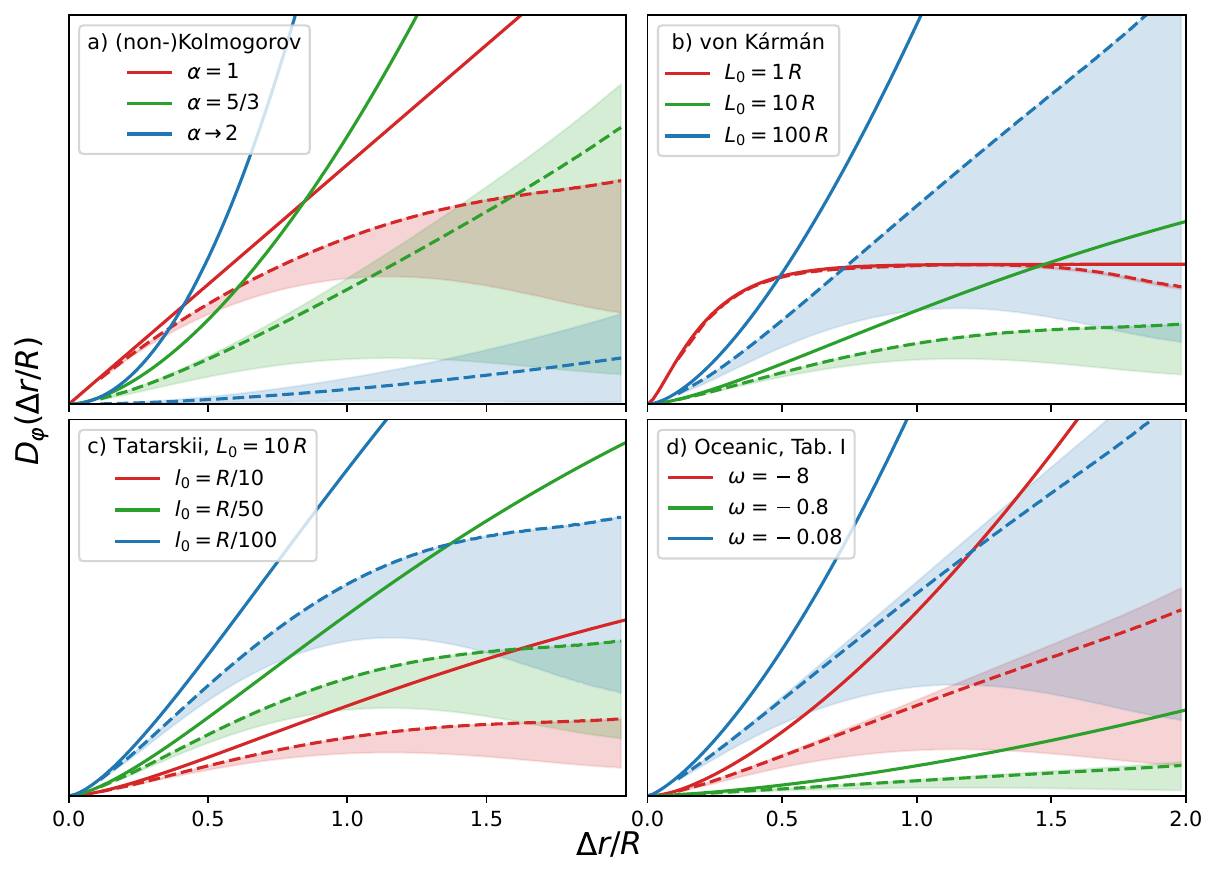}
    \caption{Rescaled structure function $D_\varphi(\Delta r/R)$ of Fourier phase screens with $0,5$ and $10$ subharmonic levels for the (non-)Kolmogorov, von Kármán, Tatarskii and oceanic spectrum,  for three different choices of the relevant spectral parameters (see legend), obtained after ensemble averaging over
    1000 realizations. The theoretically expected structure functions from \eqref{eq:spec2struc} are plotted as solid lines. The dashed lines indicate the performance of Fourier phase screens with 5 subharmonic levels and the bounds of the accompanying shaded areas correspond to plain Fourier screens (no subharmonics) and to 10 subharmonic levels, respectively, where the bound closer to the theoretical curve corresponds in all cases to the larger number of considered subharmonic levels.
    }
    \label{fig:fourier_dphi}
\end{figure}

Figure~\ref{fig:fourier_dphi} 
provides
the evaluation 
of the structure function $D_\varphi(\Delta r/R)$.
First of all, we observe in almost all cases a significant discrepancy between the theoretical
and 
computed
structure function.
The deviations become especially pronounced 
with increasing 
ratio $\Delta r/R$.
The only case for which 
the numerical
structure function matches the theory within acceptable tolerance, i.e., with relative errors of the order of one percent (see also our discussion in Sec.~\ref{sec:psm}.\ref{subsec:comparison} below) is when the outer scale of turbulence (here for the von Kármán spectrum) is 
of
the order of the aperture radius [red curve in Fig.~\ref{fig:fourier_dphi}(b)].
This behavior is precisely due to the undersampling of low-order turbulent contributions that exceed the length scale of the aperture.
Moreover, we note that, even with subharmonics, the achievable improvement of the phase screens' structure function saturates quickly with the number of considered subharmonics.
In fact, for almost all cases the screens with only five and with ten subharmonics perform almost identically to each other. Indeed, dashed curves, which correspond to five subharmonics, match the upper bounds of the shaded areas, which correspond to ten subharmonics (see Fig.~\ref{fig:fourier_dphi}).

\begin{figure}[h]
    \centering
    \includegraphics[width=\columnwidth]{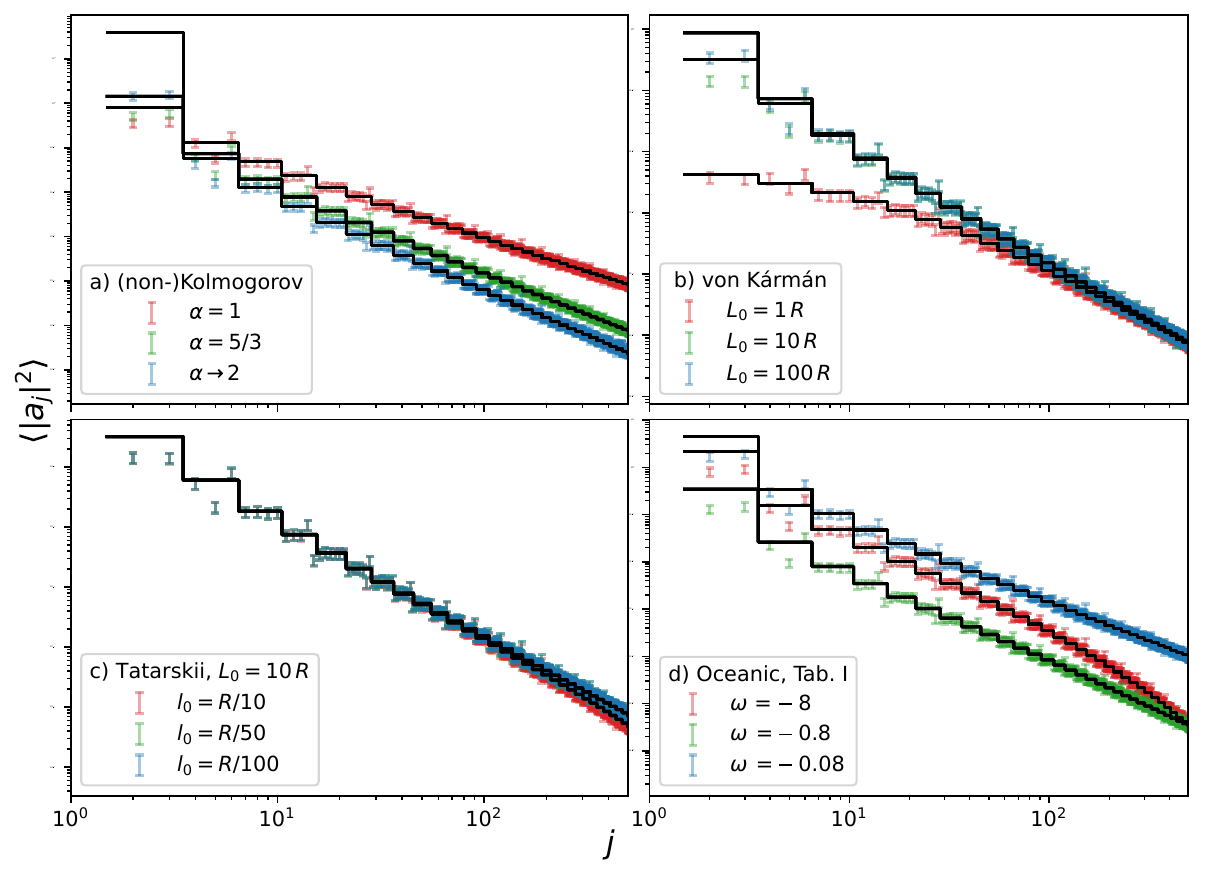}
    \caption{Log-log plot of the variances $\langle|a_j|^2\rangle$ of the Zernike expansion coefficients in \eqref{eq:zernike_exp}, for Fourier phase screens with $10$ subharmonic levels for the four different spectra given in Eqs.~(\ref{eq:kol_phase}), (\ref{eq:kar_spec}), (\ref{eq:tat_spec}) and (\ref{eq:oce_spec}) (see legend). The solid black lines connect the values theoretically expected from \eqref{eq:ajaj}. Error bars correspond to one standard deviation after ensemble averaging
    1000 disorder realizations of the turbulent medium.
    }
    \label{fig:fourier_ajs}
\end{figure}

This behavior is further confirmed by the decomposition of the phase screens with ten subharmonic levels into Zernike polynomials. As shown in Fig.~\ref{fig:fourier_ajs}, the low-order contributions do only match in the case of suppressed long-range correlations, i.e., for the von Kármán spectrum with $L_0=R$ [red points (b)].
On the other hand, high-order contributions match, as expected, perfectly with the theoretical predictions.

All in all, we thus conclude that---even with subharmonic corrections---Fourier phase screens significantly underestimate the phase structure function, which may lead to overoptimistic conclusions about the turbulence effects on the propagated beams. Only in the limiting case when long-range correlations are suppressed, i.e. when the outer scales of turbulence reach the size of the apertures, does the Fourier method produce accurate phase screens.

\subsection{Hybrid phase screens}
\label{sec:hybrid}
\label{subsec:hd}

As seen in the two previous sections, both Zernike and Fourier phase screens exhibit limitations regarding the faithful representation of a given power spectrum and structure function.
Zernike phase screens are very accurate, but the 
finite number of considered polynomials leads to a systematic underrepresentation of small scale features.
As for Fourier screens, 
they have the opposite problem: Finite grid spacing leads to a significant underestimation of large scale effects.
It is hence suggestive that a suitable combination of both methods can overcome these issues.
Such hybrid phase screens were used in Ref.~\cite{Bachmann23} for accurate simulations of dynamical Kolmogorov turbulence, and examined for atmospheric spectra when restricting the Zernike corrections to tip-tilt in Ref.~\cite{Wijerathna23}.
We here generalize these results to arbitrary power spectra and arbitrary orders of Zernike corrections.

\subsubsection{Combination of Zernike and Fourier phase screens}
The basic idea is to add a Zernike correction to an existing Fourier phase screen with aperture radius $R$ \cite{Zhu15,Wijerathna23}.
Therefore, we start by generating a plain Fourier phase screen $\varphi^{\text{FT}}(R \rho, \theta)$ (i.e. without subharmonics) by means of \eqref{eq:filter}.
Theoretically, this Fourier screen may be decomposed by means of \eqref{eq:zernike_exp} into infinitely many Zernike polynomials $Z_j(\rho,\theta)$.
In practice, however, we are restricted to a finite number $J$ of Zernike expansion coefficients (as discussed previously in Sec.~\ref{sec:psm}.\ref{subsec:zk}).
Hence, we have
\begin{align}
    \varphi^{\text{FT}}(R \rho, \theta) &= \sum_{j=1}^{\infty} a_j^{\text{FT}}\, Z_j(\rho, \theta)\nonumber\\
    &= \sum_{j=1}^{J} a_j^{\text{FT}}\, Z_j(\rho, \theta) + \sum_{j=J+1}^{\infty} a_j^{\text{FT}}\, Z_j(\rho, \theta),
    \label{eq:FTexpansion}
\end{align}
where the second term (on the second line) contains the remaining Zernike orders up to infinity.
From our analysis of Fourier phase screens in Sec.~\ref{sec:psm}.\ref{subsec:ft}, we know that the expansion coefficients $a_j^\text{FT}$, in \eqref{eq:FTexpansion} above, are deficient in representing the long-range correlations associated with lower Zernike orders of $j\leq J$, see Fig.~\ref{fig:fourier_ajs}.
This deficiency can be easily overcome by generating a finite number $J$ of accurate Zernike expansion coefficients, as described in Sec.~\ref{sec:psm}.\ref{subsec:zk}.
In other words, we generate a Zernike phase screen with $j\leq J$, that is
\begin{align}
    \varphi^{\text{ZK}}(R \rho, \theta) &= \sum_{j=1}^{J} a_j^{\text{ZK}}\,Z_j(\rho, \theta).
    \label{eq:ZKexpansion}
\end{align}
Especially for $J\lesssim 30$ of considered Zernike polynomials, \eqref{eq:ZKexpansion} can be  efficiently computed (cf. Sec. \ref{sec:psm}.\ref{subsec:comparison} and Fig.~\ref{fig:runtime} below).
Subsequently, we eliminate the deficient low-order expansion coefficients $a_j^\text{FT}$ for $j\leq J$ of the Fourier screen, i.e. the first term in \eqref{eq:FTexpansion}.
Thus, we modify the Zernike phase screen in \eqref{eq:ZKexpansion} by subtracting this term to yield a \emph{modified} Zernike phase screen
\begin{align}
    \tilde{\varphi}^{\text{ZK}}(R \rho, \theta) &= \sum_{j=1}^{J} \left(a_j^{\text{ZK}} - a_j^{\text{FT}}\right)\, Z_j(\rho, \theta).
    \label{eq:zernike_part}
\end{align}
Finally, we simply add Eqs.~(\ref{eq:FTexpansion}) and (\ref{eq:zernike_part}) to obtain our hybrid phase screen
\begin{align}
    \varphi^{\text{HD}}(R\rho, \theta) &= \varphi^{\text{FT}}(R\rho, \theta) + \tilde{\varphi}^{\text{ZK}}(R\rho, \theta)\nonumber\\
    &= \sum_{j=1}^{J} a_j^{\text{ZK}} Z_j(\rho,\theta) + \sum_{j=J+1}^\infty a_j^{\text{FT}} Z_j(\rho,\theta),
    \label{eq:HDphase}
\end{align}
which features, due to \eqref{eq:ZKexpansion}, correct long-range contributions ($j\leq J$), as well as an accurate fine structure ($j>J$) originating from \eqref{eq:FTexpansion}. A sample plot of such a \emph{hybrid} (HD) phase screen is provided in Fig.~\ref{fig:phasescreens}(c). It is noteworthy that the low-order tip and tilt contributions, featuring the Zernike screen in Fig.~\ref{fig:phasescreens}(a), are now combined with the finer structure of the Fourier screen in Fig.~\ref{fig:phasescreens}(b).

However, the Zernike expansion coefficients are \emph{not} statistically independent, as seen from the nonvanishing off-diagonal entries in Fig.~\ref{fig:ajaj}. Therefore, the merging of the independently generated constituent Fourier and Zernike phase screens according to Eqs.~(\ref{eq:zernike_part}) and (\ref{eq:HDphase}) neglects the corresponding cross correlations, introducing a small systematic error, which is discussed in more detail in Sec.~\ref{sec:appendix}, Appendix~\ref{subsec:error}.
Overall this error is very small in all cases and may safely be ignored---in accordance with the perfectly matching structure functions in Fig.~\ref{fig:hybrid_dphi}.
Moreover, this systematic error may be avoided by performing the decompositions in Eqs.~(\ref{eq:FTexpansion})-(\ref{eq:HDphase}) in the diagonal Karhunen-Loève (KL) basis instead of Zernike polynomials, cf. Sec.~\ref{sec:psm}.\ref{sssec:generationZ}.
The resulting hybrid KL screens will not suffer from this systematic error, but are more complicated to implement, since the shape of the KL polynomials depends on the underlying power spectrum.
Our preliminary results show that there is no significant advantage of KL-based hybrid screens over the here presented hybrid screens which rely on a Zernike decomposition.

\subsubsection{Statistical validation of hybrid phase screens}
Just as before (see Secs.~\ref{sec:psm}.\ref{subsec:zk} and \ref{subsec:ft}), we benchmark our hybrid screens with their structure function and investigate their Zernike expansion coefficients.
We follow the same procedure as in Sec.~\ref{sec:psm}.\ref{sssec:procedure}:
We ensemble average a set of
1000 hybrid phase screens for each of the three chosen values of the characteristic parameters of each of the previously introduced power spectra, as summarized in Tab.~\ref{tab:spec_paras} in Sec.~\ref{sec:appendix}, Appendix \ref{subsec:sim_paras}.

\begin{figure}[h]
    \centering
    \includegraphics[width=\columnwidth]{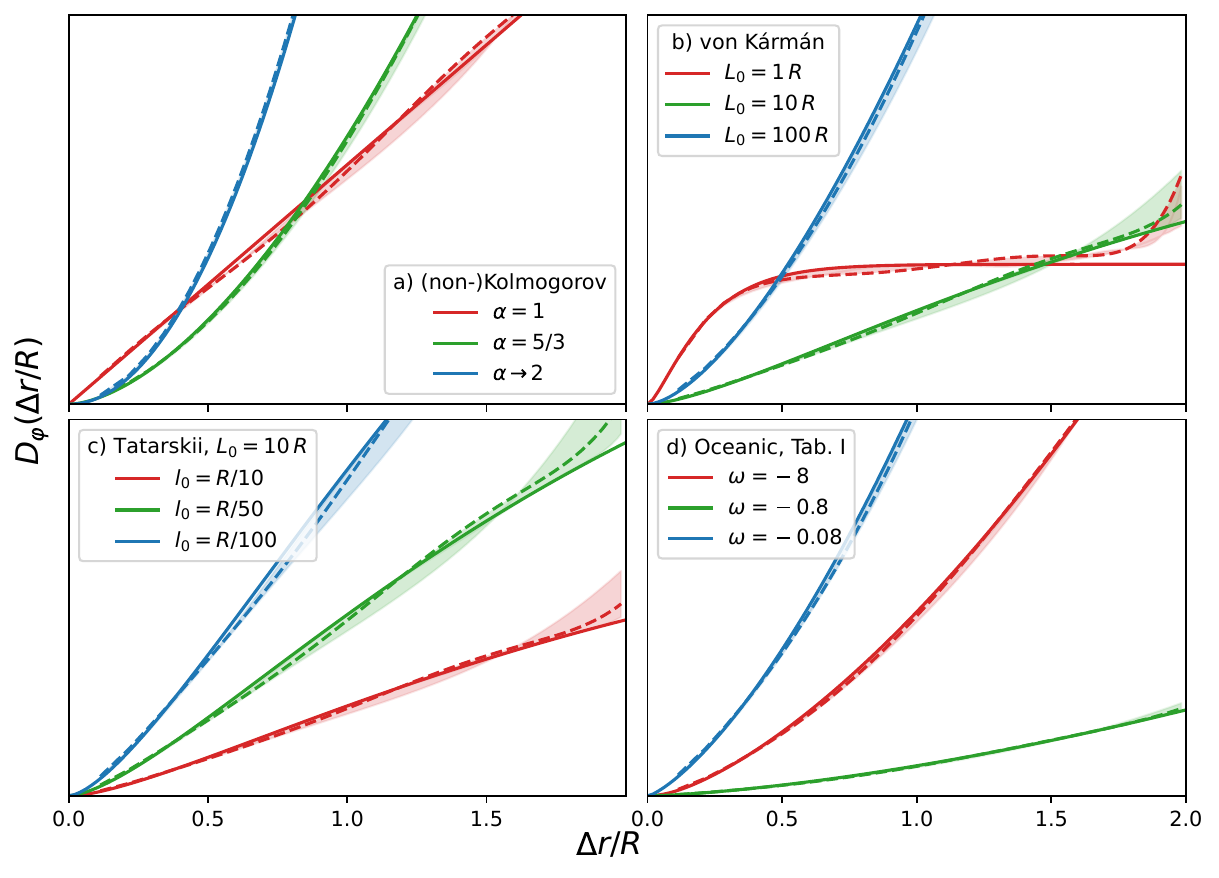}
    \caption{Rescaled structure function $D_\varphi(\Delta r/R)$ of hybrid phase screens, with $J=3,10,21$ the upper limit in \eqref{eq:zernike_exp}, for the (non-)Kolmogorov, von Kármán, Tatarskii and oceanic spectrum, for three different choices of the relevant spectral parameters (see legend), obtained after ensemble averaging over
    1000 disorder realizations of the turbulent medium. The theoretically expected structure functions from \eqref{eq:spec2struc} are plotted as solid lines. The dashed lines indicate the performance of hybrid phase screens with $J=10$, and the bounds of the accompanying shaded areas correspond to $J=3$ and $J=21$, respectively, where the bound closer to the theoretical curve corresponds, in all cases, to the larger number of considered Zernike expansion coefficients.
    }
    \label{fig:hybrid_dphi}
\end{figure}

The evaluation of the phase screens' structure function $D_\varphi(\Delta r/R)$ is shown in Fig.~\ref{fig:hybrid_dphi}.
Although we considered significantly lower numbers of Zernike corrections, i.e., $J=3,10,21$, (as compared 
to the plain Zernike screens in Sec.~\ref{sec:psm}.\ref{subsec:zk}), the structure functions show the best overall match with the theoretical expectations and the offsets that were present for plain Zernike phase screens have been successfully corrected.
In particular, for small outer scales, i.e. $L_0=R$ [red curve in Fig.~\ref{fig:hybrid_dphi}(b)], we achieve a far better agreement with theory (for further comparison also see Fig.~\ref{fig:relative_error} in Sec.~\ref{sec:psm}.\ref{subsec:comparison}).
In some cases, we observe for separations
close to the aperture diameter, i.e., $\Delta r = 2R$, a small systematic overestimation of the structure function\footnote{This effect is attributed to the limited number of considered Zernike polynomials.
As seen in Fig.~\ref{fig:zernike_pyramid}, with increasing Noll index $j$, the non-zero values of Zernike polynomials concentrate around the aperture boundary, i.e., at ratios $\Delta r/R \to 2$.
Therefore, for short outer scales of turbulence, $L_0\lesssim 10 R$, the relatively small number of considered Zernike polynomials for our hybrid screens is unable to properly compensate for contributions of lower-order Zernike polynomials (such as tip and tilt) whose non-zero values are distributed within the entire aperture, resulting in the slight systematic overestimation of the structure function as compared to theory, for $\Delta r/R \to 2$.
}.
However, this hardly affects the simulated propagation of light in practice, since the propagated beams are typically focused on an area that is significantly smaller than the aperture.

\begin{figure}[h]
    \centering
    \includegraphics[width=\columnwidth]{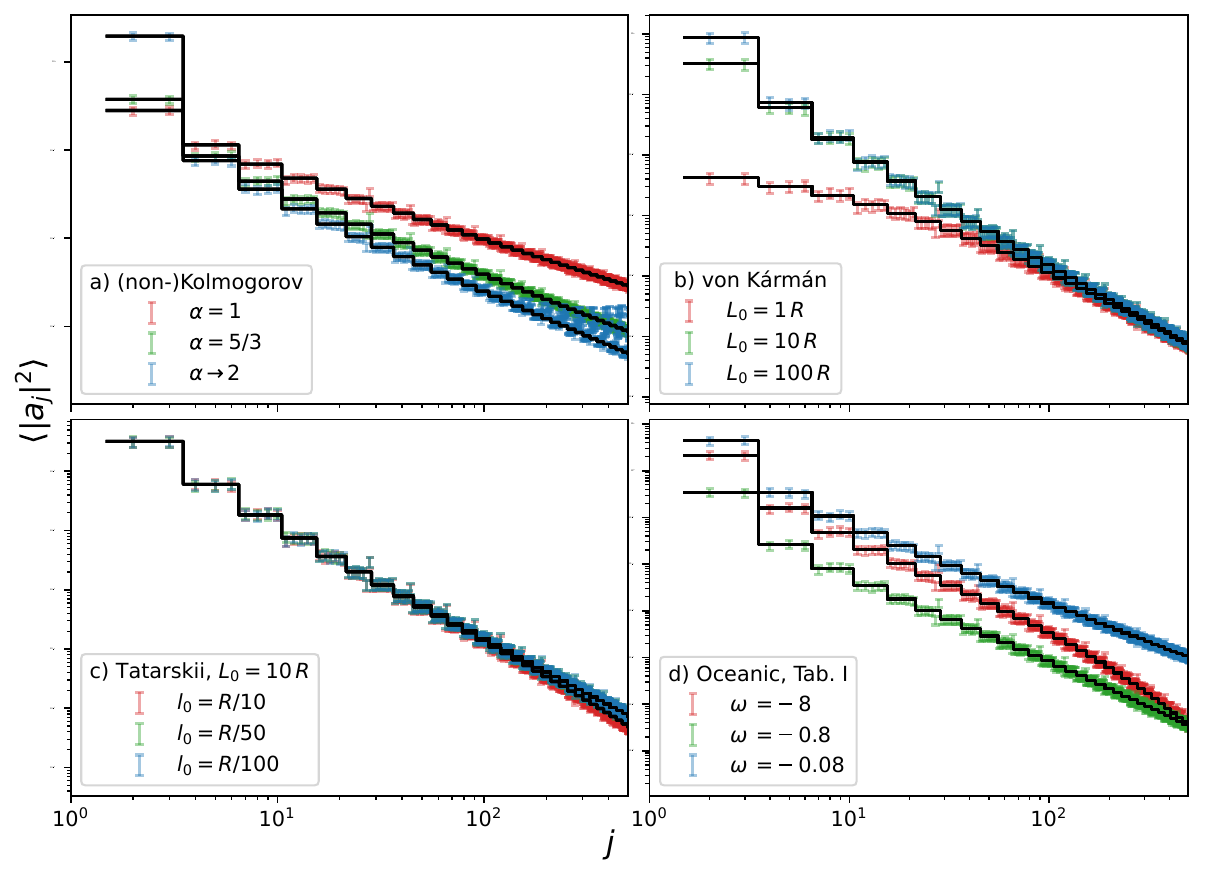}
    \caption{Log-log plot of the variances $\langle|a_j|^2\rangle$ of the Zernike expansion coefficients in \eqref{eq:zernike_exp}, for hybrid phase screens with $J=21$ Zernike corrections for the four different spectra given in Eqs.~(\ref{eq:kol_phase}), (\ref{eq:kar_spec}), (\ref{eq:tat_spec}) and (\ref{eq:oce_spec}) (see legend). The solid black lines connect the theoretically expected values from \eqref{eq:ajaj}. Error bars correspond to one standard deviation after ensemble averaging
    1000 disorder realizations of the turbulent medium.
    }
    \label{fig:hybrid_ajs}
\end{figure}

Finally, also the evaluation of the variance of the Zernike expansion coefficients gives the best overall match with theory, as depicted in Fig.~\ref{fig:hybrid_ajs}, for hybrid screens with $J=21$ Zernike corrections.
Hence, we conclude that---as expected---hybrid screens with a larger number $J$ of Zernike corrections are more accurate, and for practical implementations $J$ has to be balanced with the required computational resources as discussed in Sec.~\ref{sec:psm}.\ref{subsec:comparison} below, cf. Fig.~\ref{fig:runtime}.
The largest mismatch 
from theory, i.e., \eqref{eq:ajaj},
is found for the generalized Kolmogorov turbulence with $\alpha \to 2$ [blue points in (a)] for large $j\gtrsim 300$. We attribute this disagreement to errors of the numerical integration of \eqref{eq:ajaj}, which  increase in the vicinity of the singularity of the phase structure function at $\alpha=2$, see~\eqref{eq:nKstruc}.

\subsection{Comparison and discussion}
\label{subsec:comparison}

\begin{figure}[h]
    \centering
    \includegraphics[width=\columnwidth]{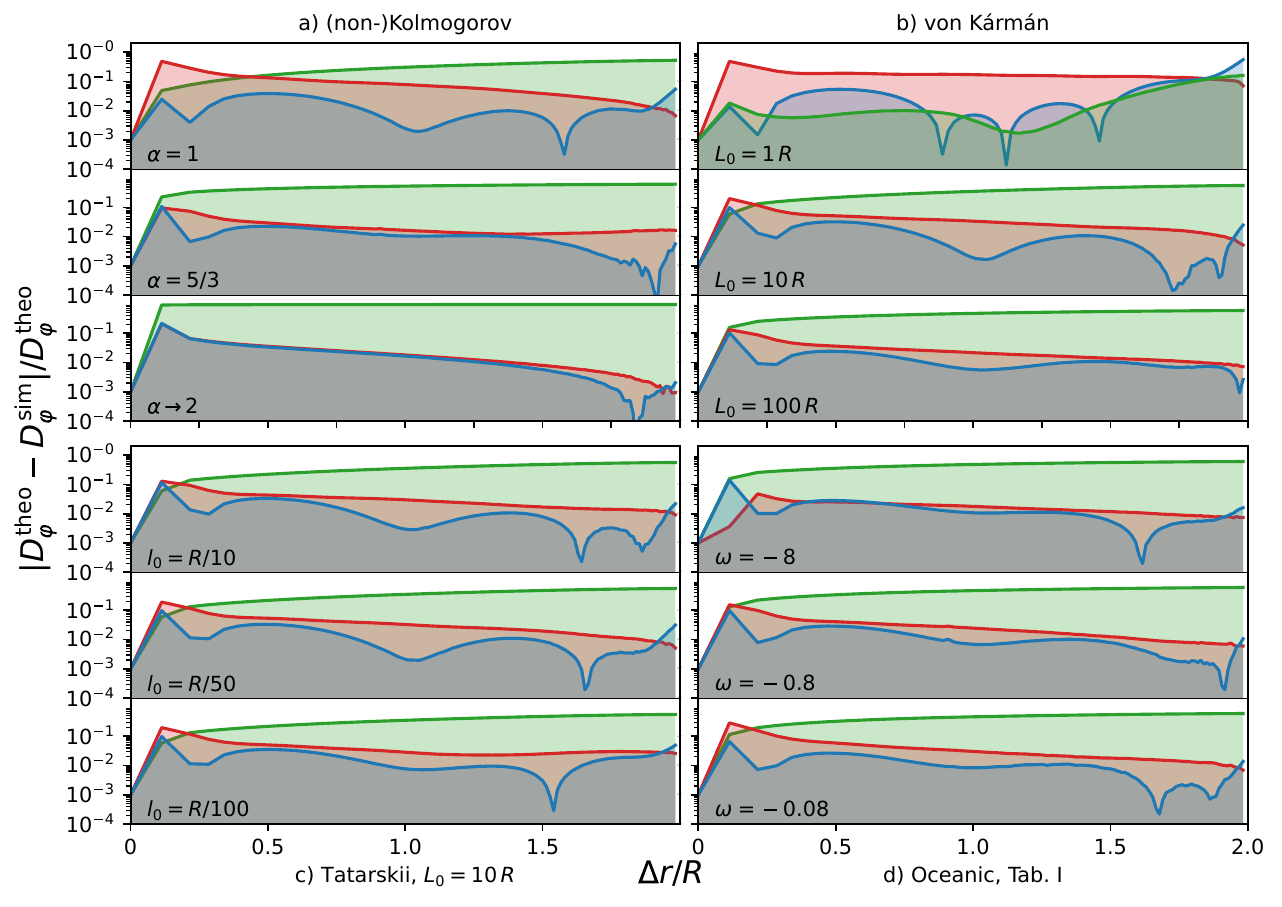}
    \caption{Relative errors of the previously obtained phase structure functions (see Figs.~\ref{fig:zernike_dphi}, \ref{fig:fourier_dphi}, \ref{fig:hybrid_dphi}) when compared to theory [see \eqref{eq:spec2struc}] of Zernike (red, $J=120$), Fourier (green, 10 subharmonics), and hybrid (blue, $J=21$) phase screens for all considered power spectra (see subcaptions and legend). The curves are ordered such that the phase screen generation methods with the smaller accumulated errors are rendered atop.
    Note that the dips in the blue curves (hybrid phase screens) correspond to crossings of the phase screens' structure function with the structure function expected from \eqref{eq:spec2struc}.
    }
    \label{fig:relative_error}
\end{figure}

Let us briefly summarize and compare the 
performance of the three different phase screen generation methods.
For comparison, we compute the relative error between their resulting structure functions (cf. Figs.~\ref{fig:zernike_dphi}, \ref{fig:fourier_dphi} and \ref{fig:hybrid_dphi}) and the theoretically expected structure function from \eqref{eq:spec2struc}, with the results shown in Fig.~\ref{fig:relative_error}.
The hybrid phase screens exhibit almost exclusively the smallest relative error, of the order of one percent, closely followed by the Zernike phase screens.
Due to the limited number $J\leq 21$ of included Zernike polynomials, the relative error of Zernike phase screens attains its maximum for small separations of $\Delta r \ll R$, and subsequently decays monotonically with increasing separations $\Delta r$.
Fourier phase screens, on the other hand, typically yield relative errors which are one or two orders of magnitude larger as compared to the other methods, and their deviation from theory grows---as expected---with increasing $\Delta r$.
The only exception is the already previously discussed case when the outer scale of turbulence is of the order of the aperture size, as seen in the top plot in Fig.~\ref{fig:relative_error}(b), for the von Kármán spectrum.
In this case, the Fourier method yields indeed the most accurate results, closely followed by the hybrid phase screens.

\begin{figure}[h]
    \centering
    \includegraphics[width=\columnwidth]{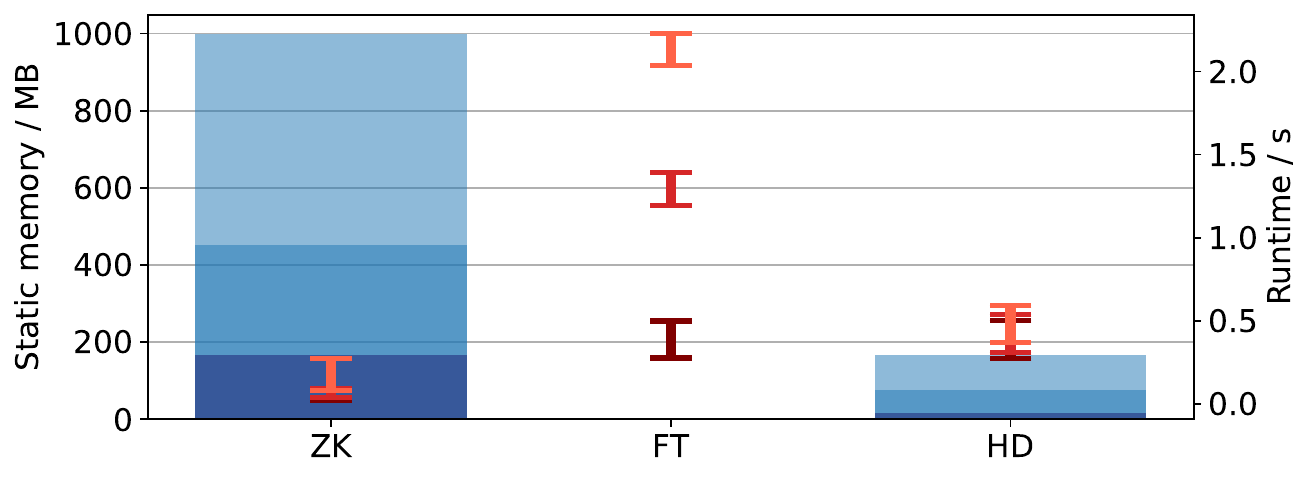}
    \caption{Average memory consumption (blue bars) and runtime (red error bars of one standard deviation) for the generation of Zernike (ZK), Fourier (FT) and hybrid (HD) phase screens, respectively. The color shading lightens with increasing number of corrections, that is for Zernike screens for $J=21, 55, 120$, for Fourier screens for $0, 5, 10$ subharmonic levels, and for hybrid screens for $J=3, 10, 21$. Both the static memory in MB and the runtime in s were determined for simulations, on an ordinary desktop machine, of phase screens with 1024x1024 pixels (the Zernike polynomials were saved as uncompressed binary containers). Naturally, the absolute runtime may vary, depending on the specific hardware configuration.
    }
    \label{fig:runtime}
\end{figure}

Hence, we conclude that Zernike phase screens are reasonably accurate
with relative errors compared to theory of the order of one percent,
even when considering only up to $J=120$ expansion coefficients [see \eqref{eq:zernike_exp}].
In fact, their accuracy may be continuously improved by
including more expansion coefficients.
This, however, would lead to 
ever longer runtimes 
for computing the needed Zernike polynomials.
Therefore, in actual applications of our method, it is sensible to compute the Zernike functions in advance, and later load them from memory.
Although this saves runtime, with increasing $J$ a non-negligible amount of static memory is needed for the phase screen generation, as shown in Fig.~\ref{fig:runtime} (blue columns on the left).
However, despite this large memory consumption, the Zernike phase screen generation takes the shortest runtime, see red error bars in Fig.~\ref{fig:runtime}, as compared to the other two methods.

In contrast, Fourier phase screens are very straightforward to implement and they do not require static memory at all.
However, as seen in Figs.~\ref{fig:fourier_dphi}, \ref{fig:fourier_ajs} and \ref{fig:relative_error}, their relatively low accuracy (relative errors when compared to theory of up to 100\,\%)
makes them suitable when the exact representation of a random medium is not crucial or when turbulent conditions reach outer scales of the order of the aperture sizes.
Although Fourier screens share the advantage that they may be generated in various geometric shapes---in contrast to the circularly constrained Zernike (and hence also hybrid) phase screens, 
their inherent undersampling of long-range correlations strongly limits their application.
Typically, long rectangular or even Fourier-based infinite screens \cite{Asse06} are employed to simulate \emph{dynamic} turbulence: Based on Taylor's hypothesis \cite{Taylor35}, the phase screens are shifted according to a transverse wind distribution.
In this case, missing long-range spatial correlations will translate into an incorrect evolution of turbulence (as demonstrated in \cite{Bachmann23}).

Additionally, the generation of Fourier screens takes significantly longer as compared to the other two methods, as seen in Fig.~\ref{fig:runtime} (red error bars in the middle column):
Even the introduction of just five subharmonic levels leads to approximately three times longer runtimes\footnote{We applied a straightforward algorithm for the generation of subharmonic Fourier phase screens based on nested loops, similar to the algorithm provided in Ref.~\cite{Schmidt10}.
At present, more optimized algorithms for phase screens generation are known ~\cite{Charnotskii:20}. In addition, the implementation in a compiled programming language, e.g., Julia or C++, as compared to interpreted languages such as Python or Matlab, may also improve the computational efficiency of Fourier screens.} as compared to Zernike or hybrid phase screens.
Hence, we conclude that the subharmonic Fourier method is neither suitable to produce accurate phase screens in terms of the structure function, nor computationally efficient.

Finally, the introduced hybrid phase screens, which combine the Zernike and Fourier method, allow for the generation of very accurate phase screens (relative errors compared to theory are typically below one percent), c.f. Figs.~\ref{fig:hybrid_dphi}, \ref{fig:hybrid_ajs} and \ref{fig:relative_error}, despite considering only a small number of Zernike polynomials.
For our purposes, we consider up to $J=21$ polynomials, which has a rather small memory footprint as compared to the full Zernike screens that require roughly six times more memory, c.f. Fig.~\ref{fig:runtime} (blue bars).
In terms of computation time, the generation of hybrid phase screens takes about twice as long as plain Zernike phase screens, but is still only insignificantly slower than the computation of plain Fourier screens (without correction by subharmonic levels), as evident from Fig.~\ref{fig:runtime} (red error bars).

\section{Conclusion}
We developed a general method for the efficient generation of highly-accurate phase screens, in order to simulate wave propagation through random continuous media.
We emphasize that our method of hybrid phase screens may be applied to an arbitrary random medium that is characterized by a phase power spectrum, provided that the wave propagation can be described by the stochastic parabolic equation.
The high accuracy of hybrid phase screens is demonstrated for four different power spectra, including non-Kolmogorov turbulence as well as the oceanic spectrum.
Moreover, for each spectrum we choose three different values of the relevant spectral parameters, to highlight diverse regimes of turbulence.
The performance of the hybrid phase screens is quantified in terms of their phase structure function, and by decomposition into Zernike polynomials and subsequent comparison to the corresponding results for Zernike and Fourier phase screens.
Our data shows that hybrid phase screens ensure the most accurate results, both in terms of their structure function and of their Zernike decomposition.
Similar to Zernike phase screens, hybrid phase screens may achieve an arbitrary level of accuracy by including more and more Zernike polynomials.
However, the computational cost in form of static memory
is approximately one order of magnitude smaller for hybrid screens.
While Fourier phase screens are straightforward to implement without any precomputations, even subharmonic corrections with runtimes several times longer than for the other two methods fail to produce
accurate
results in terms of the phase structure function.
All in all, we conclude that hybrid phase screens render the optimal choice for most applications: they provide highly accurate results while being memory efficient with decent runtimes.

\section{Acknowledgements}
The authors acknowledge fruitful discussions with Szymon Gładysz, as well as financial support by the Studienstiftung des deutschen Volkes (DB),
and support by the state of Baden-Württemberg through bwHPC and the German Research Foundation (DFG) through Grant
No. INST 40/575-1 FUGG (JUSTUS 2 cluster).
V.S. and A.B. acknowledge partial funding and support through the Strategiefonds der Albert-Ludwigs-Universität Freiburg and the Georg H. Endress Stiftung. M.I. is funded by the ANR JCJC project NoRdiC (ANR-21-CE47-0005).
This work was carried out during the tenure of an ERCIM `Alain Bensoussan’ Fellowship Programme.

\section{Appendix}
\label{sec:appendix}

\subsection{Typical parameters for the oceanic power spectrum}
\label{subsec:oce_paras}
The constants and typical parameters for the oceanic power spectrum as given in \eqref{eq:oce_spec} are listed in Tab.~\ref{tab:oce_const} below.

\begin{table}
		\caption{Constants of the oceanic power spectrum. Typical values as given in Refs. \cite{Korotkova19,Hill78}.}
		\label{tab:oce_const}
		\centering
		\begin{tabular}{r l r}
			\toprule
             & Description & Typical value\\\midrule
             $C_0$ & Obukhov–Corrsin constant & 0.72\\
             $C_1$ & Fitting constant & 2.35\\
             $\alpha_T$ & Temp. sensitivity $/\text{K}^{-1}$ & $2.6\times10^{-4}$\\
             $\chi_T$ & Temp. diffusivity $/\text{K}^{2}\,\text{s}^{-1}$ & $10^{-6}$\\
             $l_0$ & Inner scale $\text{m}$ & $10^{-3}$\\
             $\omega$ & Mixing parameter & $-1$\\
             $\chi_n$ & Refractive index diffusivity & $\left[\alpha_T^2 \chi_T (\omega - 1)^2\right]/\omega^2$\\
             $\varepsilon$ & Energy dissipation $/\text{m}^2\,\text{s}^{-3}$ & $10^{-4}$\\
             $Pr_T$ & Temp. Prandtl number & 7\\
             $Pr_S$ & Salinity Prandtl number & 700\\
             $A_T$ & Temperature decay const. & $C_0/(C_1^2\,Pr_T)$\\
             $A_S$ & Salinity decay constant & $C_0/(C_1^2\,Pr_S)$\\
             $A_{TS}$ & Temp.-Sal. decay const. & $C_0/(2C_1^2)\left[Pr_T^{-1} + Pr_S^{-1}\right]$\\
		 \bottomrule
		\end{tabular}
	\end{table}

\subsection{Analytic expressions for the covariance of Zernike expansion coefficients}
\label{subsec:aj_ana}

As mentioned in the main text, \eqref{eq:ajaj}, yielding the covariance $\langle a_j a_j^*\rangle$ of Zernike expansion coefficients, may be solved analytically for the Kolmogorov power spectrum \cite{Noll76, Wang:78, Roddier90}, see \eqref{eq:kol_phase}, as well as for the von Kármán power spectrum \cite{Conan08,Wijerathna23}, see \eqref{eq:kar_spec}.
The analytic results are provided in the following.
For the other presented spectra, the covariance of the Zernike coefficients has to be evaluated numerically.

\subsubsection{Zernike covariance for the Kolmogorov spectrum}

Substituting the Kolmogorov phase spectrum, i.e., \eqref{eq:kol_phase}, into \eqref{eq:ajaj} yields
the result
\begin{align}
    \langle a_j a_{j'}^*\rangle^{\text{Kol}} &= C_\varphi (-1)^{(n+n'-2m)/2}\,
    \sqrt{(n+1)(n'+1)}\,\nonumber\\ &\times\frac{\pi\,\Gamma\left(\frac{14}{3}\right)\Gamma\left(\frac{n+n'-\frac{5}{3}}{2}\right)}{\Gamma\left(\frac{n-n'+\frac{17}{3}}{2}\right)\,\Gamma\left(\frac{n'-n+\frac{17}{3}}{2}\right)\, \Gamma\left(\frac{n + n' +\frac{23}{3}}{2}\right)}\nonumber\\
    &\times\left(\frac{R}{2r_0}\right)^{5/3}\,\tilde{\delta}_{jj'}\,,
    \label{eq:ajaj_kol}
\end{align}
where $C_\varphi$ is given in \eqref{eq:cphi}, and the logical delta symbol $\tilde{\delta}_{jj'}$ is defined in \eqref{eq:tdelta}.
We note that this results coincides with that of Ref. \cite{Wijerathna23}, and a factor of $\pi/ 2^{5/3}$ as compared to Roddier's result in Ref. \cite{Roddier90} arises due to the angular integration, and that $R$ here represents the aperture's radius instead of its diameter. 

\subsubsection{Zernike covariance for the von Kármán spectrum}
Similarly, we may evaluate \eqref{eq:ajaj} for the von Kármán phase spectrum given by \eqref{eq:kar_spec}. We find
    \begin{align}
		\langle a_j a_{j'}^*\rangle^{\text{Kar}}&=\frac{C_\varphi\,(-1)^{(n + n'-2m)/2}\,2^{-n - n'-2}}{3 \kappa_0^{5/3}\Gamma\left(\frac{11}{6}\right)}
		\sqrt{(1 + n)(1 + n')} \pi^2\nonumber\\ &\times\csc\left[\frac{\pi}{6}\left(1 + 3n + 3n'\right)\right]
		\Bigg\{11 \times2^{n + n' + 1/3} (R \kappa_0)^{5/3}\nonumber\\
        &\times\Gamma\left(\frac{11}{6}\right) \Gamma\left(\frac{11}{3}\right)\,_3\tilde{F}_4\Bigg[\frac{11}{6},\frac{7}{3},\frac{17}{6};\frac{11-3n-3n'}{6},\nonumber\\
        &\frac{3n - 3n' + 17}{6},\frac{3n' -3n +17}{6},\frac{3n + 3n'+23}{6};R^2 \kappa_0^2\Bigg]\nonumber\\
		&-12\,(R \kappa_0)^{n + n'} \Gamma\left(\frac{n + n' + 2}{2}\right) \Gamma\left(n + n' + 3\right)\nonumber\\
        \times&_3\tilde{F}_4\Bigg[\frac{n+n'+2}{2},\frac{n+n'+3}{2},\frac{n+n'+4}{2};\nonumber\\
        &2 + n,\frac{3n+3n'+1}{6},n'+2,n + n'+3;R^2 \kappa_0^2\Bigg]\Bigg\}\, \tilde{\delta}_{jj'}\,,
        \label{eq:ajaj_kar}
\end{align}
\noindent where $_p\tilde{F}(a;b;z)_q:={_p}F_q(a;b;z)/(\Gamma(b_1)\dots\Gamma(b_q))$ denotes the generalized and regularized hypergeometric function \cite{Abramowitz65}.

\subsubsection{Zernike covariance for the non-Kolmogorov spectra}
Considering the generalization to non-Kolmogorov spectra given in \eqref{eq:kol_spec_alpha}, with the exponent $0<\alpha<2$, we find for the covariance of the Zernike coefficients, by means of \eqref{eq:ajaj},
\begin{align}
\langle a_j a_{j'}^*\rangle_{\alpha} &= C_\varphi^\alpha (-1)^{(n+n'-2m)/2}\,
    \sqrt{(n+1)(n'+1)}\,\nonumber\\
    &\times\frac{\pi \, \Gamma (\alpha +3)\,  \Gamma \left(\frac{n+n'-\alpha }{2}\right)}{\Gamma \left(\frac{n-n'+\alpha +4}{2}\right) \Gamma
   \left(\frac{n'-n+\alpha +4}{2}\right) \Gamma \left(\frac{n+n'+\alpha +6}{2}\right)}\nonumber\\
   &\times\left(\frac{R}{2 r_0}\right)^\alpha\,\tilde{\delta}_{jj'},
\end{align}
which reduces to \eqref{eq:ajaj_kol} for the exponent $\alpha=5/3$.

\subsection{Numerical implementation of Fourier phase screens}
\label{subsec:ft_details}

We collect further details regarding the numerical implementation of (subharmonic) Fourier phase screens.

\subsubsection{Fast Fourier transform (FFT) for spectral filtering}

In the numerical implementation of \eqref{eq:filter}, one has to be careful since typical FFT implementations
are based on plain spatial frequencies $\boldsymbol{f}=(f_x, f_y)$ and $f=|\boldsymbol{f}|$, instead of angular frequencies $\boldsymbol{\kappa} = 2\pi\,\boldsymbol{f}$. Performing the corresponding substitution of variables in \eqref{eq:filter}, one obtains
\begin{equation}
    \varphi(\boldsymbol{ \rho}) = \left(2\pi\right)^2\,\int d^2\boldsymbol{f}\,g\left(2\pi\,\boldsymbol{f}\right)\, \sqrt{\Phi_\varphi\left(2\pi\, f\right)}\, e^{i 2\pi\,\boldsymbol{f}\cdot\boldsymbol{\rho}},
    \label{eq:filterf}
\end{equation}
where the prefactor $\left(2\pi\right)^2$ arises due to the change of the integration variable \cite{Johansson94}.

\subsubsection{Padding of the numerical grid}
A finite grid size for the numerical implementation of the Fast Fourier Transform (FFT) introduces resonances which are most manifest at the grid boundaries. Hence, it is advisable to generate Fourier phase screens on a larger grid which is subsequently truncated to the actual size needed for the propagation \cite{Schmidt10}.
Theoretically, the grid size and its resolution should be chosen such that the pixel size is smaller than the inner scale $l_0$, while the grid's dimensions exceed the outer scale $L_0$.
In practice, however, this is not possible due to computational constraints.
In our case, we chose for all simulations a 1024x1024 grid, in which the actual 256x256 phase screen is centered.
For example, for an outer scale of $L_0=100\,R$, we would thus need a 25600x25600 grid which easily exceeds commonly available computational resources.

\subsection{Simulation parameters}
\label{subsec:sim_paras}
In the following, we summarize all relevant simulation parameters that were used to produce our data. The plotted structure functions, i.e., Figs.~\ref{fig:zernike_dphi}, \ref{fig:fourier_dphi}, \ref{fig:hybrid_dphi}, their errors in Fig.~\ref{fig:relative_error}, as well as the variances of the Zernike coefficients, i.e., Figs.~\ref{fig:zernike_ajs}, \ref{fig:fourier_ajs}, \ref{fig:hybrid_ajs}, are obtained after an ensemble average of
1000 realizations of disorder. For each phase screen model, we considered three correction levels as explained in Tab.~\ref{tab:corr_levels} below.

\begin{table}[h]
		\caption{Correction levels for the three introduced phase screen models. For Fourier phase screens, we consider three different numbers of subharmonics, where zero subharmonics correspond to a plain Fourier screen without corrections. For Zernike and hybrid phase screens, we vary the maximal number $J$ of considered Zernike polynomials which was chosen to always match a full radial order described by the radial index $n$ (c.f. in Fig.~\ref{fig:zernike_pyramid}).}
		\label{tab:corr_levels}
		\centering
		\begin{tabular}{r r r}
			\toprule
            Zernike& Fourier & $\quad$Hybrid\\
            $J,n$ & $\quad$subharmonics & $J,n$\\\midrule
            $21, 5$ & 0 & $3, 1$\\
            $55, 9$ & 5 &$10, 3$\\
            $120, 14$ & 10 & $21, 5$\\
		 \bottomrule
		\end{tabular}
\end{table}

Furthermore, for each introduced power spectrum, we choose three different spectral parameters for our simulations describing different regimes of the corresponding turbulence as summarized in Tab.~\ref{tab:spec_paras} below.
Moreover, we isolate for each spectrum the parameter that is responsible for its distinct features:
For (non-)Kolmogorov turbulence, we vary the exponent $\alpha$. In case of the von Kármán spectrum, three different outer scales $L_0$ (relative to the aperture radius $R$) were chosen. For the Tatarskii spectrum, we fixed the outer scale to ten aperture radii, and varied the inner scale $l_0$ (again relative to the aperture radius $R$). And finally, for the oceanic spectrum we kept an inner scale of $l_0=R/10$, and considered three different mixing parameters $\omega$ corresponding to either dominant salinity $(\omega=-0.08)$ or temperature $(w=-8)$ contributions, or approximately equal sharing $(\omega=-0.8)$.
Furthermore, the Fried parameter $r_0$ was factored out for all considered spectra, cf. Eqs.~(\ref{eq:kol_phase}), (\ref{eq:kar_spec}), (\ref{eq:tat_spec}),  (\ref{eq:kol_spec_alpha}), (\ref{eq:oce_spec}), in order to provide results agnostic to a specific turbulence strength $w$ (typically quantified by either the ratio with the beam waist $w=w_0/r_0$ or with the aperture size $w=R/r_0$ \cite{Andrews05}).
Hence, the choice of a specific Fried parameter simply rescales the $y$-axes of Figs.~\ref{fig:zernike_dphi}, \ref{fig:fourier_dphi}, \ref{fig:hybrid_dphi} (structure functions) and Figs.~\ref{fig:zernike_ajs}, \ref{fig:fourier_ajs}, \ref{fig:hybrid_ajs} (Zernike expansion variances).

\begin{table}[h]
		\caption{Varied parameters for each power spectrum: For each of the previously introduced spectra, we choose to evaluate our phase screen models for three instances of a given parameter.  }
		\label{tab:spec_paras}
		\centering
		\begin{tabular}{r r r r}
			\toprule
            (non-) & $\quad$von Kármán & Tatarskii & Oceanic \\
            Kolmogorov& & $\quad$$L_0=10\,R$ & $\quad$$l_0=R/10$\\
            $\alpha$ & $L_0$ & $l_0$ & $\omega$ \\\midrule
            1 & $1\,R$  & $R/100$ & $-0.08$\\
            5/3 &$10\,R$ & $R/50$ & $-0.8$\\
            2 & $100\,R$ & $R/10$& $-8$\\
		 \bottomrule
		\end{tabular}
\end{table}

\subsection{Error due to neglecting of correlations}
\label{subsec:error}

Our hybrid phase screens rely on combining a Zernike phase screen with an independently generated Fourier phase screen.
As described in \eqref{eq:FTexpansion}, we decompose the Fourier screen into Zernike polynomials and correct its deficient first $J$ orders with the expansion coefficients from the Zernike screen, see \eqref{eq:HDphase}.
Since the covariance matrix of the Zernike polynomials is not diagonal (see Fig.~\ref{fig:ajaj}), the independent generation of the two constituent phase screens and their merging in the Zernike basis introduces a systematic error by neglecting the covariances $\langle a_j\, a_{j'}^*\rangle$ of the Zernike expansion coefficients for $j>J$ and $j' \leq J$ (and vice versa), as illustrated in Fig.~\ref{fig:error_sketch} below.

\begin{figure}[h]
    \centering
    \includegraphics[width=0.7\columnwidth]{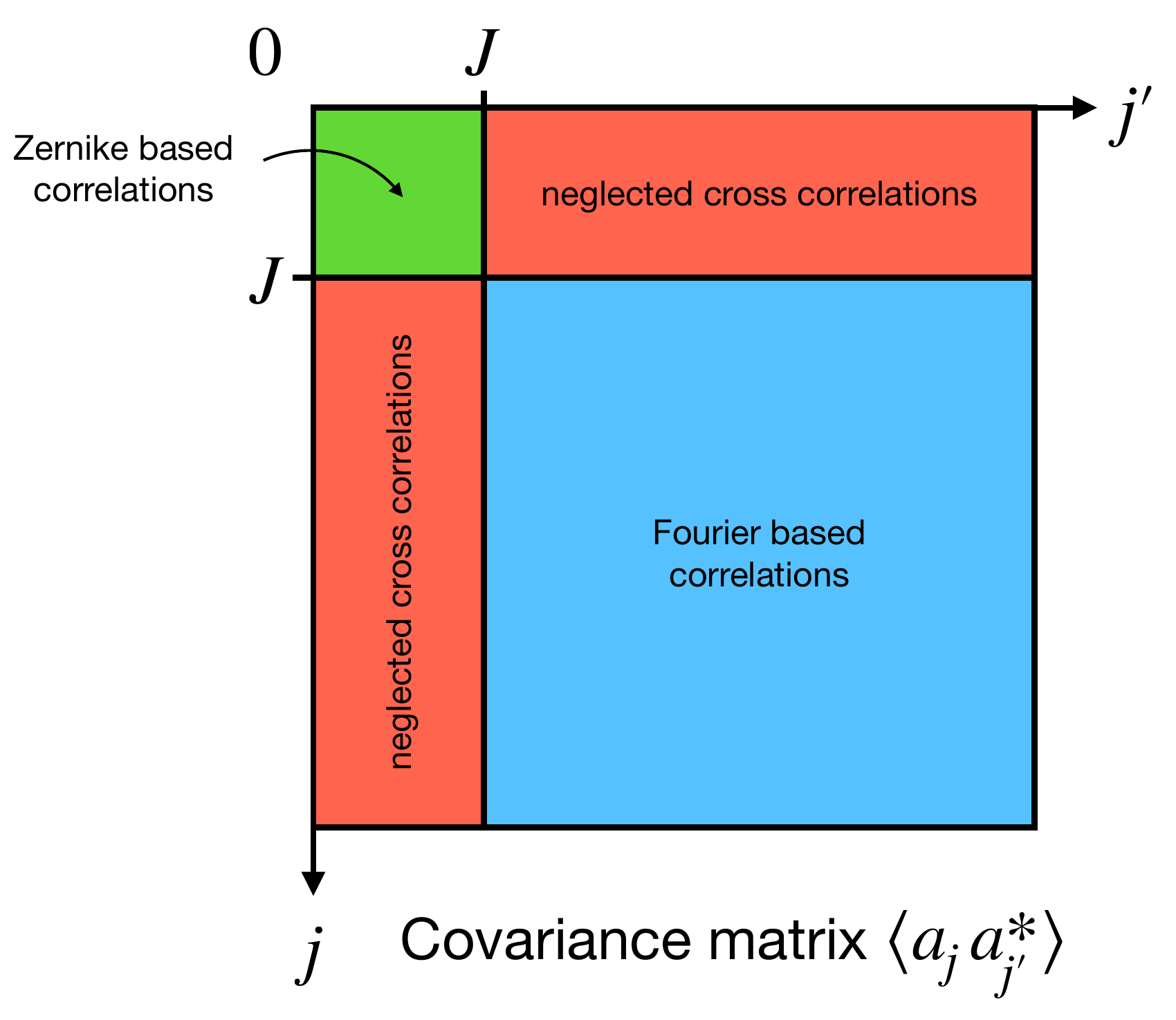}
    \caption{
    Sketch of the constituents of the Zernike covariance matrix of a hybrid phase screen.
    The green region corresponds to corrected correlations by means of the Zernike phase screen and the blue region represents the  correlations due to the constituent Fourier phase screen.
    The red areas feature systematic errors due to the independent generation of the constituent Zernike and Fourier phase screens.
    }
    \label{fig:error_sketch}
\end{figure}

\begin{figure}[h]
    \centering
    \includegraphics[width=0.7\columnwidth]{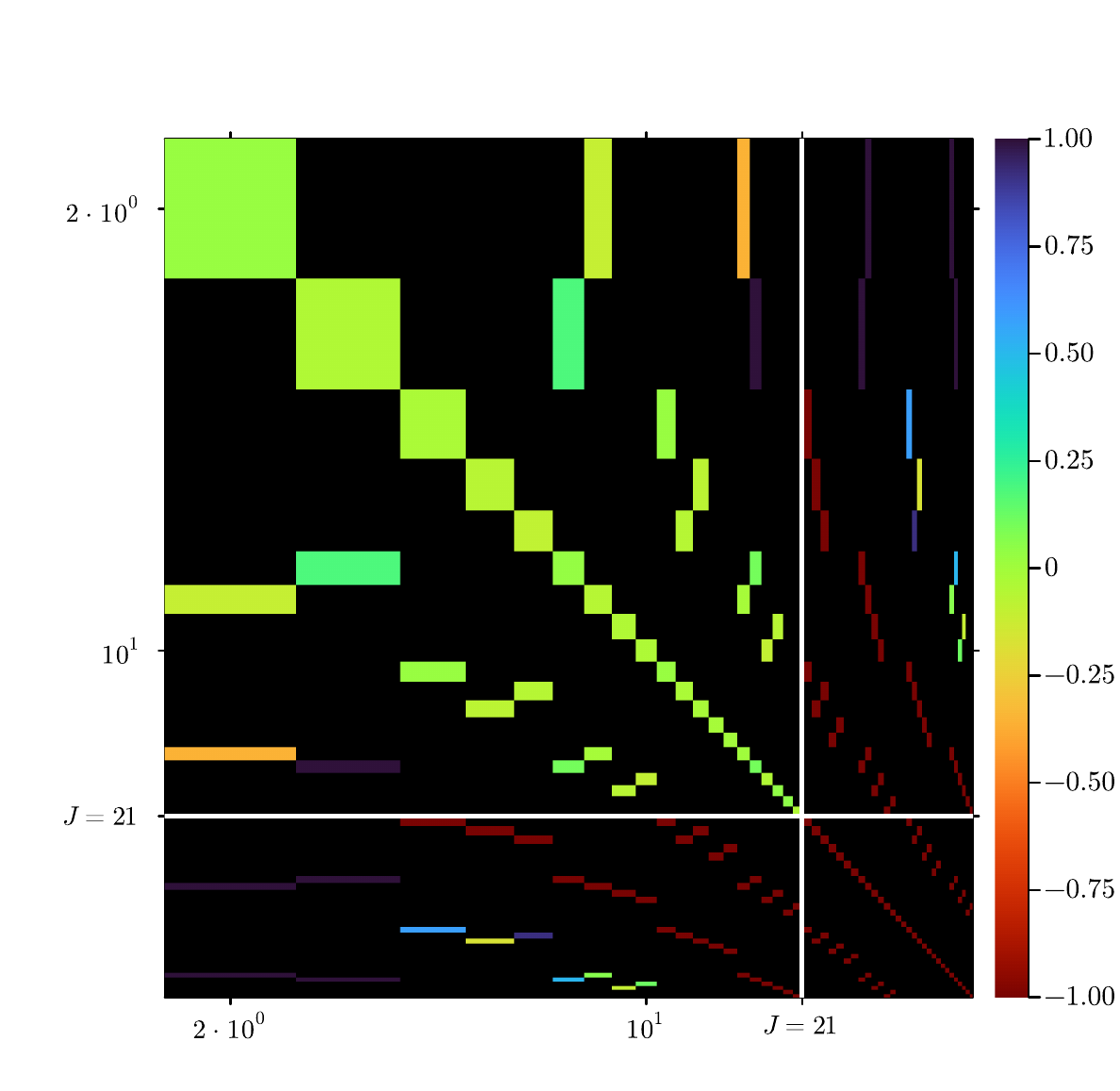}
    \caption{
    Relative error $\log|\langle a_j a_{j'}^*\rangle / \langle a_j a_{j'}^*\rangle_\text{theo}|$ of the Zernike covariances obtained from the ensemble average of 1000 Zernike phase screens with $J=21$ included Zernike orders for Kolmogorov turbulence $(\alpha=5/3)$.
    }
    \label{fig:covmat-zk}
\end{figure}

\begin{figure}[h]
    \centering
    \includegraphics[width=0.7\columnwidth]{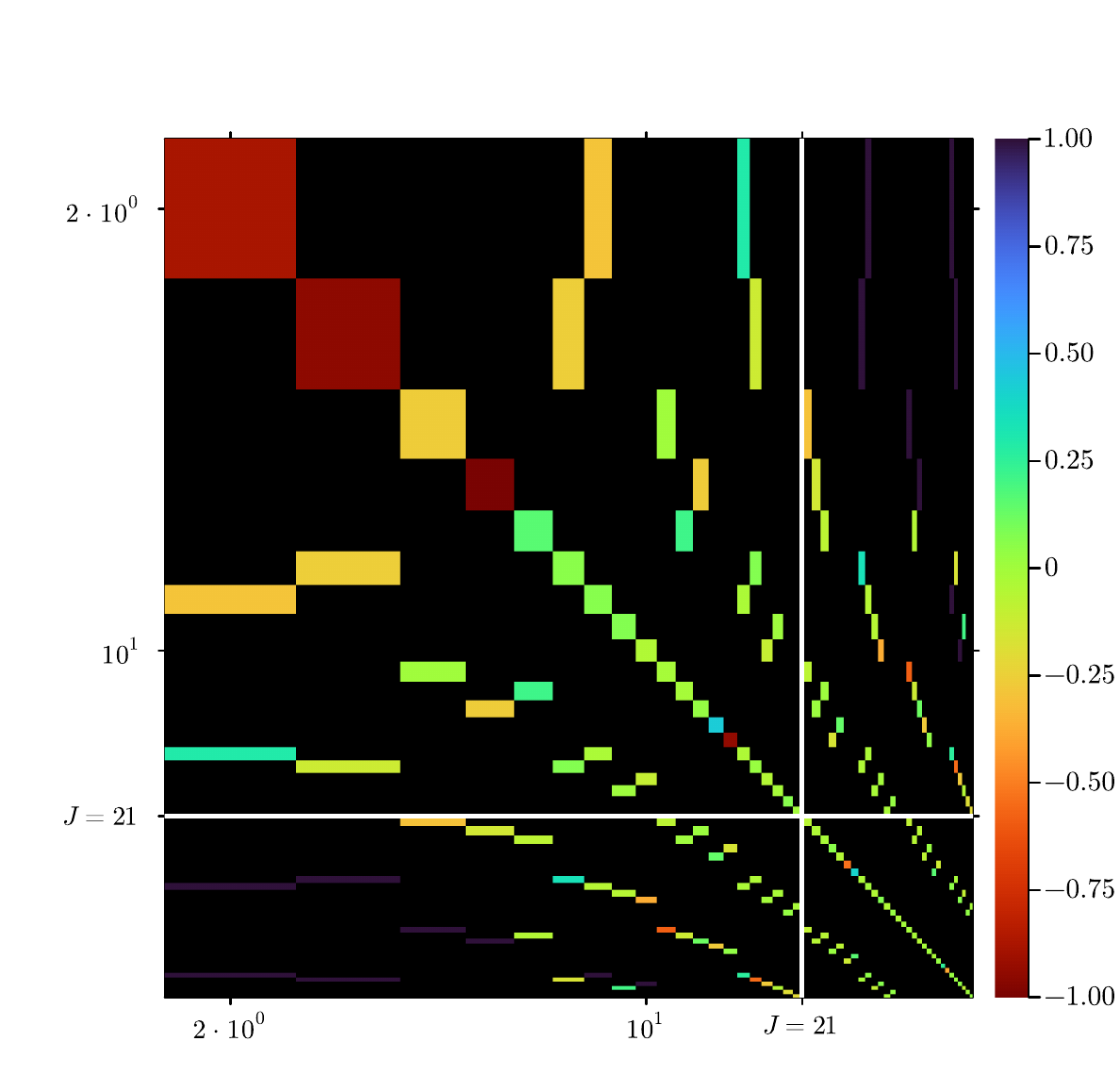}
    \caption{
   Same as Fig.~\ref{fig:covmat-zk}, but obtained from the ensemble average of 1000 subharmonic Fourier phase screens with 10 subharmonics for Kolmogorov turbulence $(\alpha=5/3)$.
    }
    \label{fig:covmat-ft}
\end{figure}

\begin{figure}[h]
    \centering
    \includegraphics[width=0.7\columnwidth]{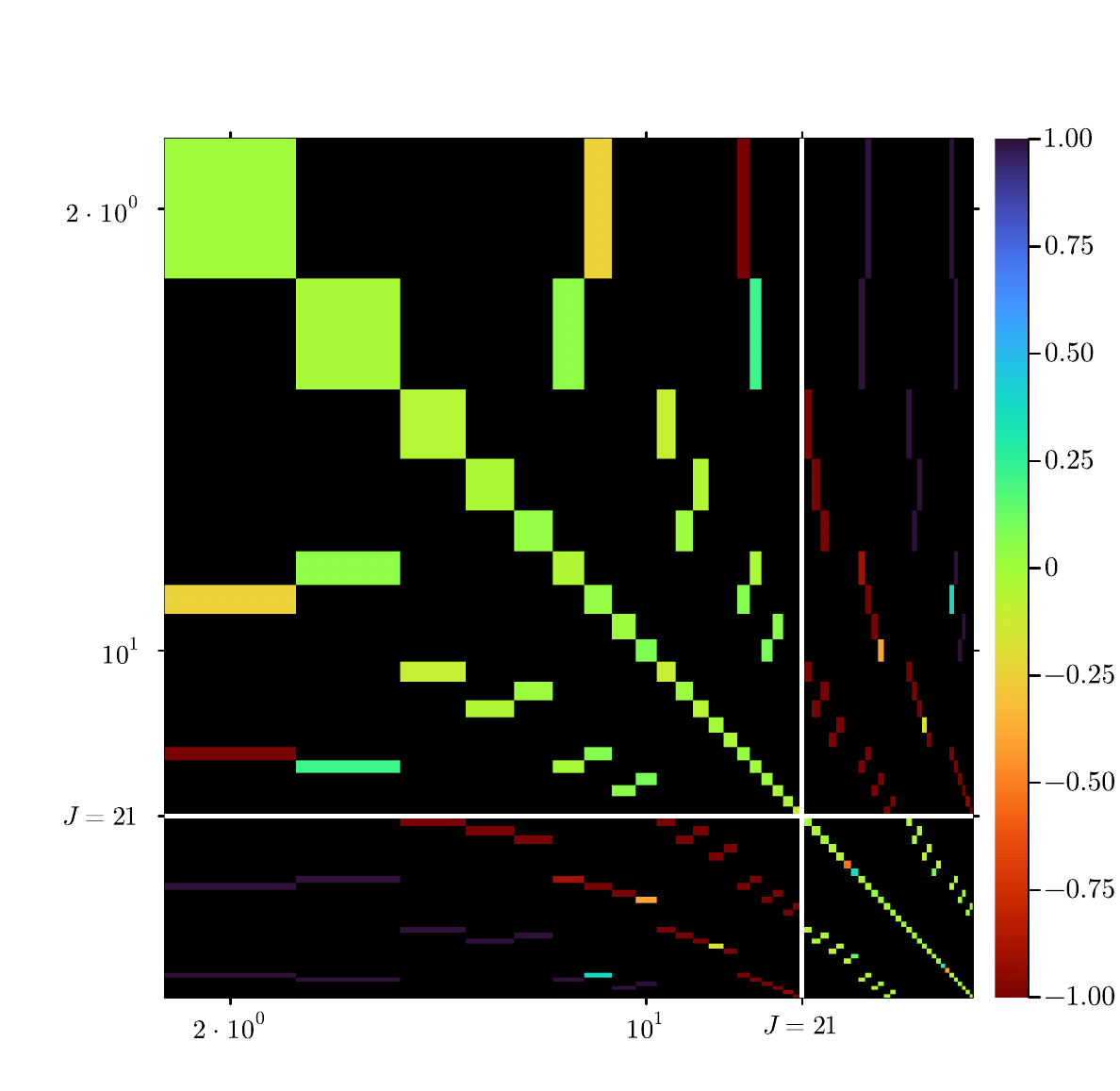}
    \caption{
    Same as Fig.~\ref{fig:covmat-zk}, but obtained from the ensemble average of 1000 hybrid phase screens with $J=21$ correction orders for Kolmogorov turbulence $(\alpha=5/3)$.
    }
    \label{fig:covmat-hd}
\end{figure}

Figures~\ref{fig:covmat-zk}-\ref{fig:covmat-hd} plot the relative error of the Zernike covariance matrices compared to theory from \eqref{eq:ajaj} for Zernike, subharmonic Fourier, and hybrid phase screens.
For the Zernike phase screen with $J=21$ Zernike orders, we observe---as expected---for $j,j'\leq J$ very good agreement with theory, as shown in Fig.~\ref{fig:covmat-zk}, where the main discrepancy is found for offdiagonal elements with large $j$ or $j'$ close to $J$, probably due to the limited sample size of 1000 realizations.
Furthermore, it is clearly visible that the Zernike covariances are missing for $j>J$ or $j'>J$.

For the subharmonic Fourier phase screens, we observe---in accordance with our findings in Fig.~\ref{fig:fourier_ajs}---that the error of the Zernike covariances is greatest for small $j$ or $j'$, as shown in Fig.~\ref{fig:covmat-ft}.
In particular, the low order Zernike contributions such as tip-tilt and astigmatism show relative errors of around one order of magnitude, while covariances corresponding to higher $j$ and $j'$ are overall matching well with theory.

Finally, Fig.~\ref{fig:covmat-hd} plots the Zernike covariances as obtained for hybrid phase screens with $J=21$ Zernike orders.
By construction, we obtain very good agreement with theory for $j,j' \leq J$, matching the behavior of Zernike phase screens as shown in Fig.~\ref{fig:covmat-zk}.
Additionally, now for $j$ and $j'> J$, we find as well accurate Zernike covariances, originating from the constituent Fourier phase screen, cf. Fig.~\ref{fig:covmat-ft}.
However, the regions with $j>J$ and $j'\leq J$ (and vice versa) exhibit the previously discussed systematic error, where we observe the expected mismatch from theory due to the neglected cross correlations. 

\bibliography{singularmodes4}

\end{document}